\documentclass[twocolumn,preprintnumbers, amssymb,amsmath,aps,prd,floatfix,nofootinbib,superscriptaddress,showpacs,nolongbibliography]{revtex4-2}

\usepackage{amsmath}
\usepackage{amssymb}
\usepackage{graphicx}
\graphicspath{{figures/}{paper/figures/}}
\usepackage[normalem]{ulem}
\usepackage{xcolor}
\usepackage{hyperref}

\begin{document}

\title{Tracing Gluon Saturation through Hadronization at EIC}

\author{Weiyao Ke}
\affiliation{Key Laboratory of Quark \& Lepton Physics (MOE) and Institute of Particle Physics, Central China Normal University, Wuhan 430079, China}
\affiliation{Southern Center for Nuclear-Science Theory (SCNT), Institute of Modern Physics, Chinese Academy of Sciences, Huizhou, Guangdong 516000, China}

\author{Xu Liu}
\affiliation{Key Laboratory of Particle Physics and Particle Irradiation (MOE), Institute of Frontier and Interdisciplinary Science, Shandong University, Qingdao, Shandong 266237, China}

\author{Yu Shi}
\affiliation{Key Laboratory of Particle Physics and Particle Irradiation (MOE), Institute of Frontier and Interdisciplinary Science, Shandong University, Qingdao, Shandong 266237, China}
\affiliation{CPHT, CNRS, \'Ecole Polytechnique, Institut Polytechnique de Paris, 91120 Palaiseau, France}

\author{Xin-Nian Wang}
\affiliation{Key Laboratory of Quark \& Lepton Physics (MOE) and Institute of Particle Physics, Central China Normal University, Wuhan 430079, China}

\author{Jian Zhou}
\affiliation{Key Laboratory of Particle Physics and Particle Irradiation (MOE), Institute of Frontier and Interdisciplinary Science, Shandong University, Qingdao, Shandong 266237, China}
\affiliation{Southern Center for Nuclear-Science Theory (SCNT), Institute of Modern Physics, Chinese Academy of Sciences, Huizhou, Guangdong 516000, China}

\begin{abstract}
Gluon saturation provides a window into the nonlinear nature of the strong interaction in nuclear matter. One direct consequence of saturation  is the  $\boldsymbol{k}_T$ broadening  in the final state.
We investigate how hadronization reshapes conventional signatures of gluon saturation at EIC within a complete event-generator framework. To this end, we implement in eHIJING an initial-state-radiation algorithm based on nonlinear small-$x$ evolution and complete the events with beam remnants, final-state radiation, and hadronization. In our simulations, the signal of parton-level nuclear $\boldsymbol{k}_T$ broadening is strongly diluted in both the nucleon energy correlator and leading-dihadron azimuthal decorrelation after hadronization. Global hadronic recoil, by contrast, remains sensitive to the underlying $\boldsymbol{k}_T$ broadening. We further demonstrate that Bayesian unfolding of the global hadronic recoil provides access to the underlying hard-scattering $\boldsymbol{k}_T$ distribution. These results establish the global hadronic recoil as a promising saturation observable at the EIC.
\end{abstract}

\maketitle

\noindent\textit{\textbf{Introduction.}}
    Gluon saturation is a central prediction of nonlinear QCD dynamics in the high-occupancy regime~\cite{Gribov:1983ivg,Mueller:1985wy,Mueller:1989st,McLerran:1993ni,McLerran:1993ka,McLerran:1994vd,Gelis:2010nm,Iancu:2003xm}, and its experimental confirmation is one of the primary scientific goals of the Electron--Ion Collider (EIC). The saturation scale $Q_s^2(x)$ grows with decreasing initial parton momentum fraction $x$ and increasing nuclear size $A$, making electron--nucleus collisions at the EIC an ideal process for studying the saturation phenomenon~\cite{Accardi:2012qut,AbdulKhalek:2021gbh}. The reconstructed transverse-momentum ($\boldsymbol{k}_T$) distribution of initial-state gluons, including its nuclear broadening and its dependence on the gluon momentum fraction $x$, provides one of the most direct probes of gluon saturation. Substantial theoretical progress has also been made in small-$x$ evolution and the calculation of saturation-sensitive observables~\cite{Kovchegov:1999yj,Albacete:2010sy,Dominguez:2011wm,Dominguez:2011br,Metz:2011wb,Iancu:2015joa,Dumitru:2015gaa,Altinoluk:2015dpi,Dumitru:2016jku,Boer:2016fqd,Hatta:2016dxp,Boer:2017xpy,Kotko:2017oxg,Dumitru:2018kuw,Mantysaari:2019csc,Boussarie:2019ero,Salazar:2019ncp,Iancu:2020jch,Shi:2021hwx,Bergabo:2021woe,Boussarie:2021ybe,Zhao:2021kae,Caucal:2021ent,Boer:2021upt,Hagiwara:2021xkf,Iancu:2021rup,Taels:2022tza,Bergabo:2022tcu,Bergabo:2022zhe,Caucal:2022ulg,Tong:2022zwp,Iancu:2022lcw,Hatta:2022lzj,Zhang:2021tcc,Bergabo:2023wed,Tong:2023bus,Rodriguez-Aguilar:2023ihz,Caucal:2023nci,Caucal:2023fsf,Shao:2024nor,Altinoluk:2024vgg,Caucal:2024nsb,Altinoluk:2025dwd,Caucal:2025qjg,Marquet:2025jdr,Shao:2026doo,Kutak:2011fu,Jung:2000hk,Baranov:2021uol,Hautmann:2022xuc,Lipatov:2023ypn,Marquet:2007vb,Stasto:2018rci,Mueller:2012uf,Mueller:2013wwa,Zheng:2014vka,Akcakaya:2012si,Kotko:2015ura,vanHameren:2014ala,vanHameren:2016ftb,vanHameren:2019ysa,vanHameren:2020rqt,Al-Mashad:2022zbq,Gao:2026azd,Xiao:2017yya,Zhou:2016tfe,Zhou:2018lfq,Dominguez:2011gc,Beuf:2014uia,Liu:2022xsc,Zheng:2019zul,Marquet:2005zf}.
A key phenomenological question is whether parton-level saturation signals survive parton showering, hadronization, and nuclear-remnant breakup.


At the Born level in the Breit frame, the $\boldsymbol{k}_T$ imbalance of the hard $q\bar q$ pair produced through $\gamma^* g^*\to q\bar q$ is set entirely by the $\boldsymbol{k}_T$ of the small-$x$ gluons. Parton showering, hadronization, and nuclear-remnant breakup subsequently redistribute this recoil among the final-state particles. This redistribution can dilute the nuclear broadening observed in leading-particle correlations and angular energy-flow projections. A quantitative particle-level assessment therefore requires a fully exclusive description of the final state.

In this work, we extend eHIJING~\cite{Ke:2023xeo} with a small-$x$ component that implements the nonlinear small-$x$ initial-state shower developed in Refs.~\cite{Shi:2022hee,Shi:2023ejp}, and we include final-state showers and hadronization. This fully exclusive framework allows us to determine which saturation signatures survive into the hadronic final state.

We first examine two representative semi-inclusive DIS probes of $\boldsymbol{k}_T$ broadening: leading-dihadron azimuthal decorrelation and the nucleon energy correlator (NEC). In leading-dihadron correlations, saturation-induced $\boldsymbol{k}_T$ broadening manifests itself as a wider away-side peak~\cite{Dominguez:2011wm,Zheng:2014vka}.  Additional soft radiation, however, generates Sudakov recoil that further decorrelates the pair and strongly dilutes the saturation signal. Furthermore, such decorrelation can be induced by nonperturbative hadronization.
 At the relatively low final-state energies relevant to EIC small-$x$ kinematics, a fragmentation-function treatment alone may be insufficient for a quantitative description of the hadronic final state. Jet reconstruction does not automatically circumvent this difficulty, as anti-$k_T$ dijets do not faithfully reconstruct the four-momenta of the underlying hard partons in this kinematic region.

A complementary observable is the nucleon energy correlator (NEC), an energy-weighted angular distribution in DIS~\cite{Liu:2022wop}. Its sensitivity to saturation at small-$x$ has been studied in Refs.~\cite{Liu:2023aqb,Mantysaari:2025mht,Kang:2026hig,Mantysaari:2026zte}. However, fragmentation and breakup of the nuclear remnant can contribute substantially to the energy flow in the same angular region as the predicted saturation signal. A quantitative interpretation of the NEC therefore requires explicit control of these nonperturbative contributions.

To overcome these limitations, we seek observables that are more inclusive with respect to the redistribution of transverse momentum. We introduce the global hadronic transverse recoil and characterize its event-by-event fluctuations through a two-particle transverse-momentum correlation $C_T$. A clear difference in $C_T$ between $e+\mathrm{Au}$ and $e+p$ collisions survives at the hadron level. In addition, a Bayesian unfolding of the global recoil can recover the underlying parton-level $\boldsymbol{k}_T$ distribution of the hard $q\bar q$ pair. These results demonstrate that hadronization redistributes, rather than erases, the transverse-momentum broadening generated by saturation, establishing global hadronic recoil as a promising probe of small-$x$ parton dynamics at the EIC.

\bigskip

\noindent\textit{\textbf  {Parton-Level Baseline.}}
To isolate saturation-induced broadening before final-state evolution, we first determine how the transverse-momentum imbalance of the hard $q\bar q$ pair depends on the target. For each target, the unintegrated gluon distribution $\mathcal{N}_{p/A}(\eta,k_T)$, with $\eta=\ln(x_{\mathrm{sw}}/x_g)$, is initialized at $x_{\mathrm{sw}}=0.01$ as a Gaussian in $k_T$, with a width set by the corresponding saturation scale. The distribution is then evolved toward smaller $x_g$ using the kinematically constrained folded GLR equation~\cite{Shi:2022hee,Shi:2023ejp}:
\begin{equation}
  \frac{\partial\mathcal{N}_A}{\partial\eta}
  =\mathcal{K}_{\mathrm{real}}\!\otimes\mathcal{N}_A
  -\mathcal{K}_{\mathrm{virtual}}\!\otimes\mathcal{N}_A
  -\mathcal{C}_{\mathrm{nl}} \bar \alpha_s \mathcal{N}_A^2 .
  \label{eq:folded-evolution}
\end{equation}
The real term governs resolvable emissions in the backward shower, while the virtual and nonlinear terms enter the accompanying suppression factor. For the inputs used here, the gold distribution is broader in $k_T$ than the proton distribution at fixed $x_g$.

Hard events are generated by combining the evolved gluon TMD with the off-shell $\gamma^*+g^*\to q+\bar q$ impact factor. The sampled transverse momentum $\boldsymbol{k}_T$ is the imbalance of the hard $q\bar q$ pair, $k_T=|\boldsymbol{p}_{T,q}+\boldsymbol{p}_{T,\bar q}|$~\cite{Dominguez:2011wm,Boussarie:2021ybe}. Starting from the selected incoming off-shell gluon, the backward GLR shower reconstructs the associated small-$x$ initial-state radiation (ISR) until the evolving $t$-channel gluon reaches $x_{\mathrm{sw}}=0.01$. Although the dynamics at $x>x_{\mathrm{sw}}$ can be absorbed into the boundary condition for inclusive distributions, fully exclusive event generation requires an explicit radiation history. We therefore continue the backward evolution with a DGLAP shower. The subsequent remnant construction and color assignment enforce exact event-by-event transverse-momentum conservation. Further details of the eHIJING implementation are provided in the Appendix.

Using this setup, we generate $e+p$ and $e+\mathrm{Au}$ events at identical beam energies and with the same phase-space cuts. For comparison, we also generate a PYTHIA~8.3 $e+p$ DIS sample~\cite{Bierlich:2022pfr}, restricted to events with $x_g<0.01$. Its prehadronization hard-pair distribution provides a reference based on conventional ISR.

\begin{figure}[t]
  \includegraphics[width=\columnwidth]{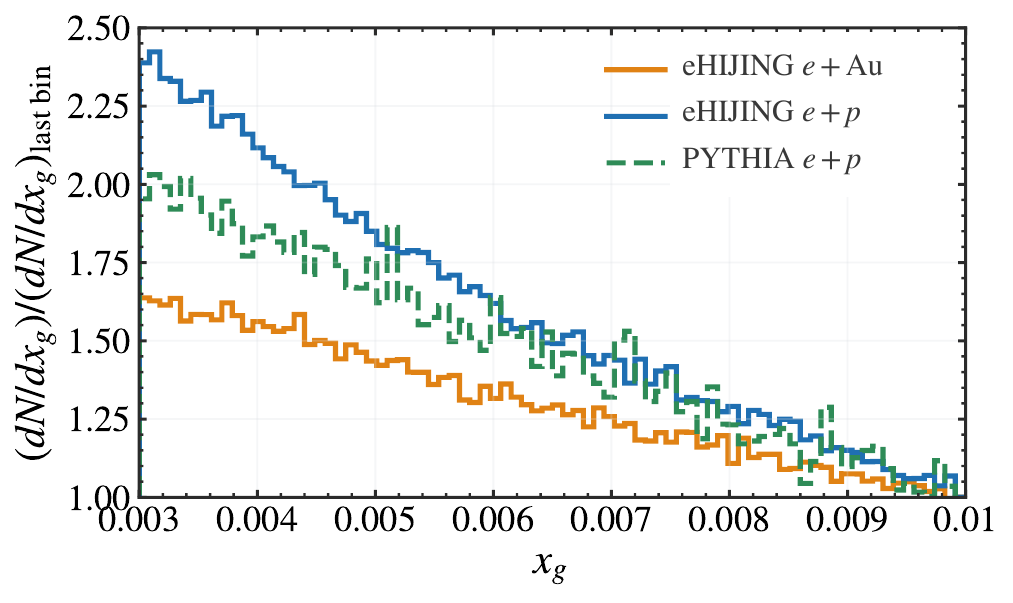}
  \caption{Distributions of the incoming-gluon momentum fraction $x_g$ for selected hard-scattering events. The eHIJING $e+\mathrm{Au}$ and $e+p$ samples are shown in orange and blue, respectively; the green dashed curve shows the PYTHIA~8.3 $e+p$ sample with conventional initial-state radiation. All samples are restricted to $0.003\leq x_g\leq0.01$, and each distribution is normalized by its value in the largest $x_g$ bin.}
  \label{fig:xg-vetoed-pythia}
\end{figure}

Figure~\ref{fig:xg-vetoed-pythia} shows the distributions of the incoming-gluon momentum fraction $x_g$ for the selected hard-scattering events. The eHIJING $e+p$ and $e+\mathrm{Au}$ samples and the PYTHIA~8.3 proton reference populate the same small-$x$ region, $0.003\leq x_g\leq0.01$. All subsequent target comparisons are restricted to this common interval.

\begin{figure}[t]
  \includegraphics[width=\columnwidth]{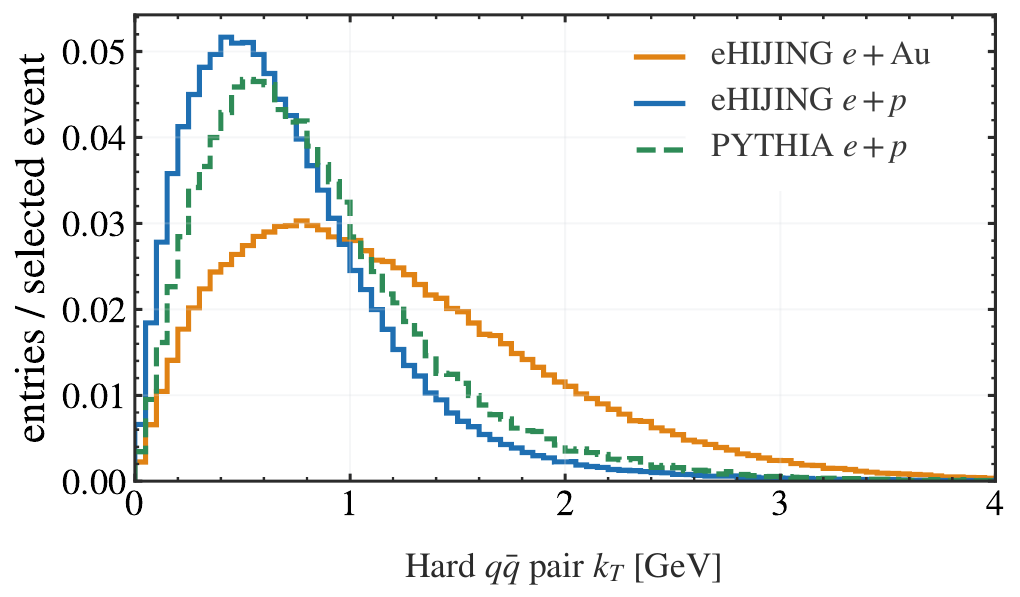}
  \caption{The transverse-momentum imbalance of the hard $q\bar q$ pair in the Breit frame. The orange and blue curves show the eHIJING generator-level $e+\mathrm{Au}$ and $e+p$ distributions, respectively, before showering and hadronization. The green dashed curve shows the PYTHIA~8.3 $e+p$ control sample, constructed from the partonic descendants obtained after the parton shower and primordial-$k_T$ assignment but before hadronization. The $e+\mathrm{Au}$ distribution is broader than the $e+p$ distribution.}
  \label{fig:hard-pair-kt-vetoed-pythia}
\end{figure}

At generator level, the transverse-momentum imbalance of the hard $q\bar q$ pair is broader in $e+\mathrm{Au}$ than in $e+p$, as shown in Figure~\ref{fig:hard-pair-kt-vetoed-pythia}. We next examine how final-state showering and hadronization transmit this initial difference to hadron-level observables.
\bigskip

\begin{figure}[t]
  \includegraphics[width=\columnwidth]{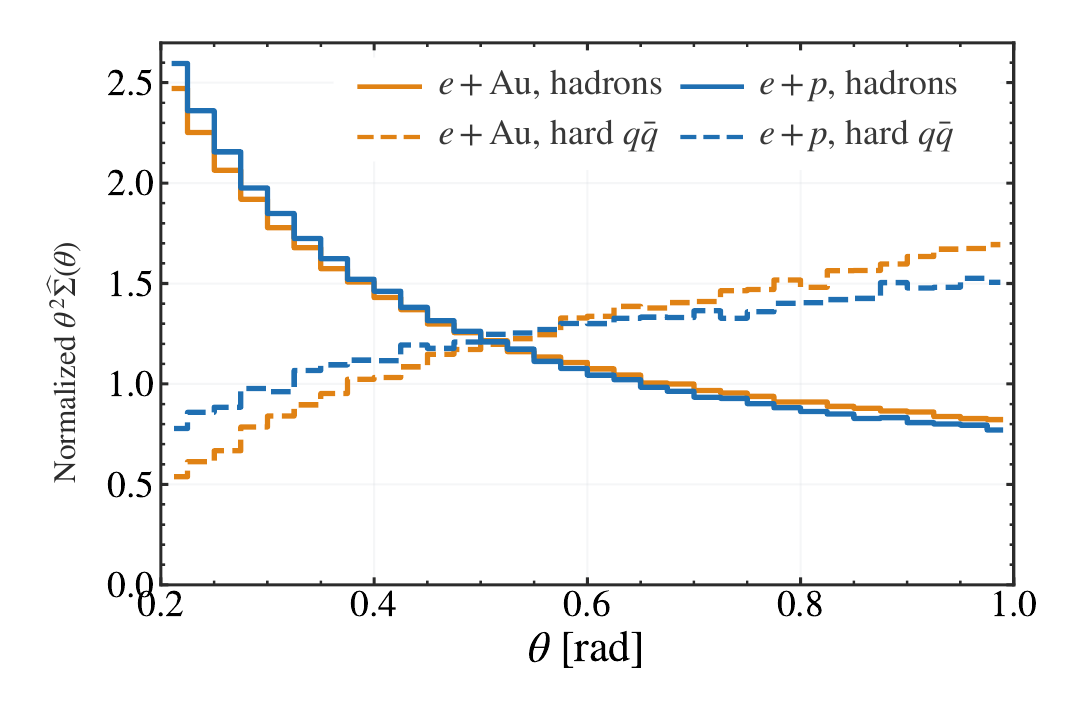}
  \caption{Distribution of the nucleon energy correlator (NEC), $\theta\widehat{\Sigma}(\theta)$, for $0.002<x_B<0.004$ and $15\leq Q^2<23\,\mathrm{GeV}^2$. Orange and blue denote $e+\mathrm{Au}$ and $e+p$; solid and dashed curves denote stable hadrons and the pre-shower hard $q\bar q$ pair, respectively. For each target, the stable-hadron and pre-shower hard-pair curves use the same normalization convention over $0.2\leq\theta_h<1.0$. The angle $\theta_h$ is measured from the incoming-hadron direction.}
  \label{fig:neec-baseline}
\end{figure}

\noindent\textit{\textbf{Hadronization Effects on Saturation Observables.}}
Following Ref.~\cite{Liu:2023aqb}, we define the event-normalized nucleon energy correlator in the $d\theta^2$ convention as
\begin{equation}
  \widehat{\Sigma}(\theta)
  =\frac{1}{N_{\mathrm{evt}}}
   \sum_{\mathrm{events}}\sum_{h\in\mathrm{event}}
   \frac{E_h}{E_N}
  \delta(\theta^2-\theta_h^2).
  \label{eq:energy-angular}
\end{equation}
Here, $E_h$ and $E_N$ are the hadron and incoming-nucleon energies, respectively, and $\theta_h$ is measured relative to the incoming-hadron direction. The sum runs over stable hadrons of all charges and species. Figure~\ref{fig:neec-baseline} shows $\theta\widehat{\Sigma}(\theta)$ for $e+p$ and $e+\mathrm{Au}$ collisions at the pre-shower hard-parton and stable-hadron levels over $0.2\leq\theta<1.0$, with all curves normalized according to Eq.~\eqref{eq:energy-angular}.

At the hard-parton level, the nuclear suppression at small $\theta$ is clearly visible. At the hadron level, however, the contribution to the small-angle NEC from the full hadronic final state is much larger than that from the hard $q\bar q$ pair alone, and the $e+p$ and $e+\mathrm{Au}$ distributions become nearly indistinguishable. The $q\bar q$ pair inherits the color-octet charge of the incoming gluon and is therefore color connected to the target remnant. Within the string-hadronization model used here, fragmentation of this extended color system, together with target-remnant fragmentation, populates the target-going small-angle region and dilutes the relative nuclear modification carried by the hard pair. Probing saturation with the NEC therefore requires quantitative control of these contributions.

\begin{figure}[t]
  \includegraphics[width=\columnwidth]{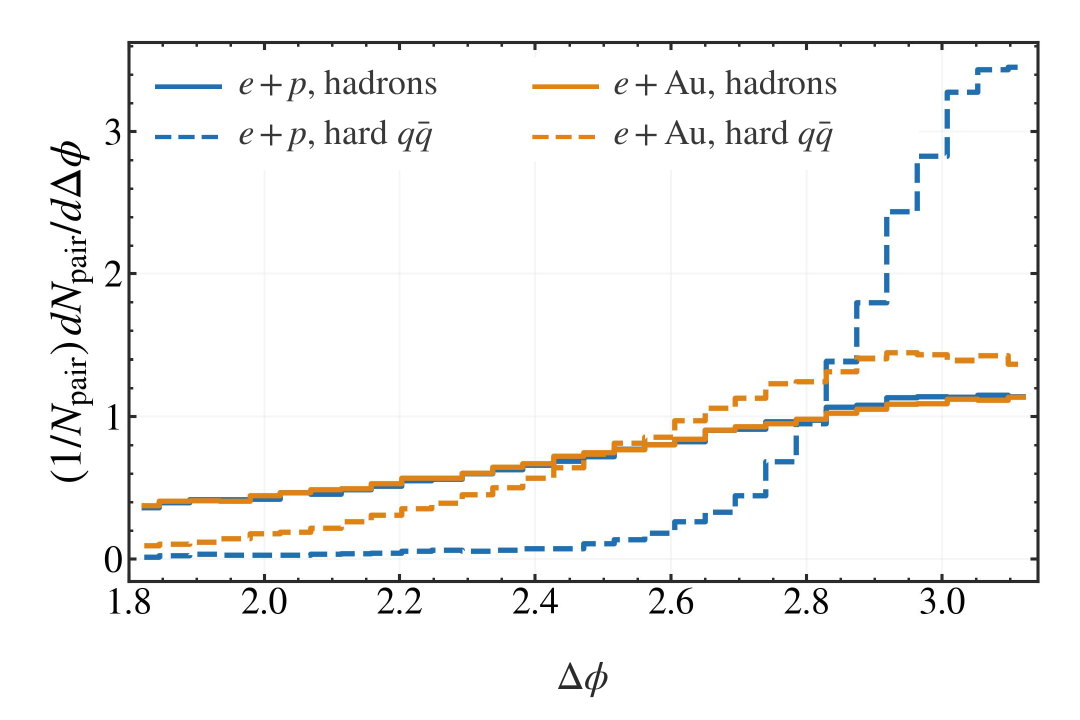}
  \caption{Leading-hadron azimuthal decorrelation in the Breit frame. Solid curves show the $e+\mathrm{Au}$ (orange) and $e+p$ (blue) distributions for the two leading stable hadrons with $p_{T,1}>1.0\,\mathrm{GeV}$ and $p_{T,2}>0.6\,\mathrm{GeV}$. No restrictions are imposed on charges or hadron species. Dashed curves show the distributions for hard $q\bar q$ pairs selected independently at the pre-shower level, with $p_{T,1}>3.0\,\mathrm{GeV}$ and $p_{T,2}>2.0\,\mathrm{GeV}$. The distributions are integrated over the selected $x_B$ range and normalized to unit area over $1.8\leq\Delta\phi\leq\pi$.}
  \label{fig:dihadron-baseline}
\end{figure}

We next study the leading-hadron away-side decorrelation as a complementary test. Figure~\ref{fig:dihadron-baseline} compares the unit-normalized away-side $\Delta\phi$ distributions for pre-shower hard $q\bar q$ pairs and for the two leading hadrons. At the Born level, the acoplanarity distribution is clearly broader in $e+\mathrm{Au}$ than in $e+p$. Showering and hadronization redistribute the recoil of the hard pair among multiple hadrons, broadening the leading-hadron distributions for both targets and largely washing out their separation. A similar dilution of the transverse-momentum broadening signal has been found in studies incorporating Sudakov resummation~\cite{Stasto:2018rci,Marquet:2025jdr,Gao:2026azd}.  

The strong loss of sensitivity in the hadron-level NEC and dihadron correlations, together with the limited fidelity of low-energy dijet reconstruction, motivates the use of global hadronic transverse recoil. Although showering and hadronization redistribute the hard-system recoil among many particles, the vector sum of these particles' transverse momenta over a broad rapidity region that excludes the far-forward target remnant can remain correlated with the initial-state $\boldsymbol{k}_T$.
\bigskip

\noindent\textit{\textbf{Global Hadronic Recoil and the Hard-Pair Imbalance.}}
Before parton showering and hadronization, the transverse-momentum imbalance $\boldsymbol{k}_T$ of the $q\bar q$ pair directly reflects the transverse momentum of the small-$x$ gluon. To improve the sensitivity in the final state, we define the global hadronic recoil as
\begin{equation}
  \boldsymbol{r}_T
  =\sum_{h:\,|y_h|<2}\boldsymbol{p}_{T,h},
  \qquad r_T=\lvert\boldsymbol{r}_T\rvert .
  \label{eq:hadron-recoil}
\end{equation}
The sum runs over all stable hadrons in the Breit-frame rapidity window $|y_h|<2$. Final-state dynamics map the $\boldsymbol{k}_T$ distribution of the hard $q\bar q$ pair onto the global-recoil $\boldsymbol{r}_T$ distribution. The vector sum preserves the directional correlations among these hadrons. The target-dependent ISR is concentrated in the target-going region. Restricting the recoil sum to hadrons with $-2<y_h<2$ therefore suppresses the direct ISR contribution while retaining sensitivity to the recoil of the hard $q\bar q$ pair.

We construct $\boldsymbol{r}_T$ for every event in the common analysis phase space. The ensemble average $\langle \boldsymbol{r}_T\rangle$ vanishes, so we investigate the second moment $\langle \boldsymbol{r}_T^2 \rangle$, with $\boldsymbol{r}_T^2=\sum_h\sum_{h'} \boldsymbol{p}_{T,h}\cdot \boldsymbol{p}_{T,h'}$ event by event. To remove the auto-correlation in $\langle \boldsymbol{r}_T^2 \rangle$, we define the two-particle transverse-momentum correlation for each target and $x_B$ bin as
\begin{align}
  C_T
  &\equiv
  \frac{\langle \boldsymbol{r}_T^2-\sum_h(p_{T,h})^2\rangle}
       {\langle \sum_h(p_{T,h})^2\rangle} =\frac{\langle \sum_{h\neq h'}\boldsymbol{p}_{T,h}\!\cdot
                    \boldsymbol{p}_{T,h'} \rangle}{\langle \sum_h(p_{T,h})^2\rangle}.
  \label{eq:ct-correlation}
\end{align}
The subtraction removes self-correlations, so $C_T$ measures the net $p_T$-weighted alignment of distinct hadron pairs. Positive (negative) values correspond to net alignment (anti-alignment) of hadron pairs within the acceptance. By construction, $C_T$ is a ratio of ensemble moments rather than an event-by-event variance.

\begin{figure}[t]
  \includegraphics[width=\columnwidth]{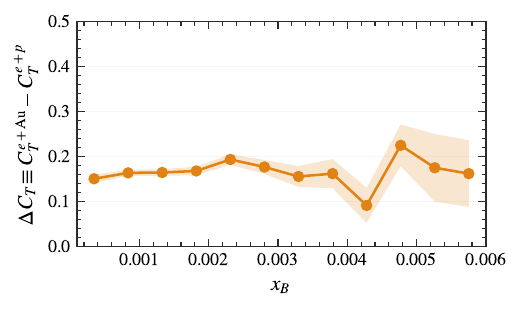}
  \caption{Hadron-level difference between targets $\Delta C_T=C_T^{e+\mathrm{Au}}-C_T^{e+p}$, as a function of $x_B$, integrated over $Q^2$. The sums entering $C_T$ are evaluated event by event using stable hadrons with $|y_h|<2$ in the Breit frame. All events within the common $(x_B,Q^2)$ phase space are included. The shaded band indicates the statistical $1\sigma$ uncertainty.}
  \label{fig:coherent-recoil-ct}
\end{figure}

At hadron level, $C_T$ remains sensitive to the target. Figure~\ref{fig:coherent-recoil-ct} shows the target difference $\Delta C_T\equiv C_T^{e+\mathrm{Au}}-C_T^{e+p}$ within $10^{-4}\leq x_B\leq0.006$. In every $x_B$ bin, $\Delta C_T$ remains positive, indicating that the net $p_T$-weighted pair contribution in gold is systematically shifted toward alignment relative to that in the proton.

To determine if this hadron-level recoil difference retains information about the initial gluon transverse momentum,
we test whether the generator-level $k_T$ distribution can be reconstructed with useful accuracy from the measured $r_T$ distribution together with the leptonic kinematics $(x_B,Q^2)$.
We denote the parton-level variables by $X=(x_g,k_T)$ and the hadron-level observables by $Y=(x_B,Q^2,r_T)$. Because the recoil is measured within a finite rapidity acceptance, $r_T$ and $k_T$ are statistically correlated rather than equal event by event. For $e+p$, this relation is encoded in the response matrix $R^p_{YX}\equiv P_p(Y| X)$,
which gives the probability for an event in partonic bin $X$ to populate hadron-level bin $Y$. The response incorporates the effects of showering, beam-remnant kinematics and color connections, hadronization, and particle-level acceptance. Including $x_g$ in $X$ is essential because the correlation between $r_T$ and $k_T$ depends on the small-$x$ evolution interval. A response constructed from $k_T$ alone would average over this dependence and could therefore be biased by differences in the sampled $x_g$ distributions.

For a given hadronization model, the proton and nuclear samples use the same models for final-state showering, hadronization, and target fragmentation. Restricting $r_T$ to central rapidity further reduces direct contributions from the target-dependent initial-state shower and remnant sector. These considerations motivate the approximation $P_{\mathrm{Au}}(Y | X)\approx P_p(Y| X)$
at fixed $(x_g,k_T)$. We therefore construct $R^p$ from simulated $e+p$ events and apply it to the hadron-level $e+\mathrm{Au}$ distribution; its explicit construction is described in Appendix~\ref{app:ibu}. The unfolded $k_T$ distribution is then compared with the directly generated $e+\mathrm{Au}$ parton-level ground truth. This cross-target closure test determines both whether $r_T$, together with $(x_B,Q^2)$, retains sufficient information about the initial gluon $k_T$ and whether the proton response can be transferred to a nuclear target.

We then use iterative Bayesian unfolding~\cite{DAgostini:1994fjx} to reconstruct the joint $e+\mathrm{Au}$ distribution $N_{\mathrm{Au}}(x_g,k_T)$ from the hadron-level distribution $N_{\mathrm{Au}}(x_B,Q^2,r_T)$. Both the response matrix and the prior are obtained entirely from simulated $e+p$ events. The hadron-level $e+\mathrm{Au}$ distribution is used as the input to the unfolding, while the corresponding parton-level distribution is reserved as the ground truth for the closure test. Integrating the reconstructed distribution over $x_g$ gives the hard-pair $k_T$ distribution. Further details of the unfolding procedure are given in Appendix~\ref{app:ibu}.

The upper panel of figure~\ref{fig:ibu-reconstruction} shows that, for four different hadronization models, the unfolded results captures the fraction of  $e+\mathrm{Au}$ events for three different $k_T$ bins. We can quantify the agreement between two normalized distributions $P$ and $Q$ using the total variation distance,
\begin{equation}
D_{\mathrm{TV}}(P,Q)
=\frac{1}{2}\sum_i |P_i-Q_i| ,
\end{equation}
where $D_{\mathrm{TV}}=0$ indicates identical distributions and $D_{\mathrm{TV}}=1$ indicates distributions with no overlap. 
The four hadronization models---PYTHIA string, HERWIG~7 cluster~\cite{Bellm:2015jjp}, AHADIC++ cluster~\cite{Winter:2003tt}, and JETSCAPE hybrid hadronization~\cite{JETSCAPE:2023ewn}---yield $D_{\mathrm{TV}}=0.068$--$0.073$ for $e+\mathrm{Au}$ after integration over $x_g$. The narrow range of $D_{\mathrm{TV}}$ values for $e+\mathrm{Au}$ shows that the proton response retains similar accuracy across all four hadronization models.
As for the joint distribution, an independent $e+p$ closure tests yield $D_{\mathrm{TV}}=0.0023$--$0.0045$.

For the PYTHIA string sample, propagating the reconstructed partonic distribution through $R^p$ reproduces the particle-level distribution. The comparison gives $D_{\mathrm{TV}}=0.031$ over the full $(x_B,Q^2,r_T)$ space. After integration over $x_B$ and $Q^2$, the corresponding value for the five-bin $r_T$ distribution is $D_{\mathrm{TV}}=0.022$. At the chosen coarse resolution, the agreement at partonic and particle levels establishes global hadronic recoil as a means to reconstruct the transverse-momentum imbalance of the hard $q\bar q$ pair in $e+\mathrm{Au}$ after hadronization. These results also provide a benchmark for cross-model and detector-level studies.

\begin{figure}[t]
  \includegraphics[width=\columnwidth]{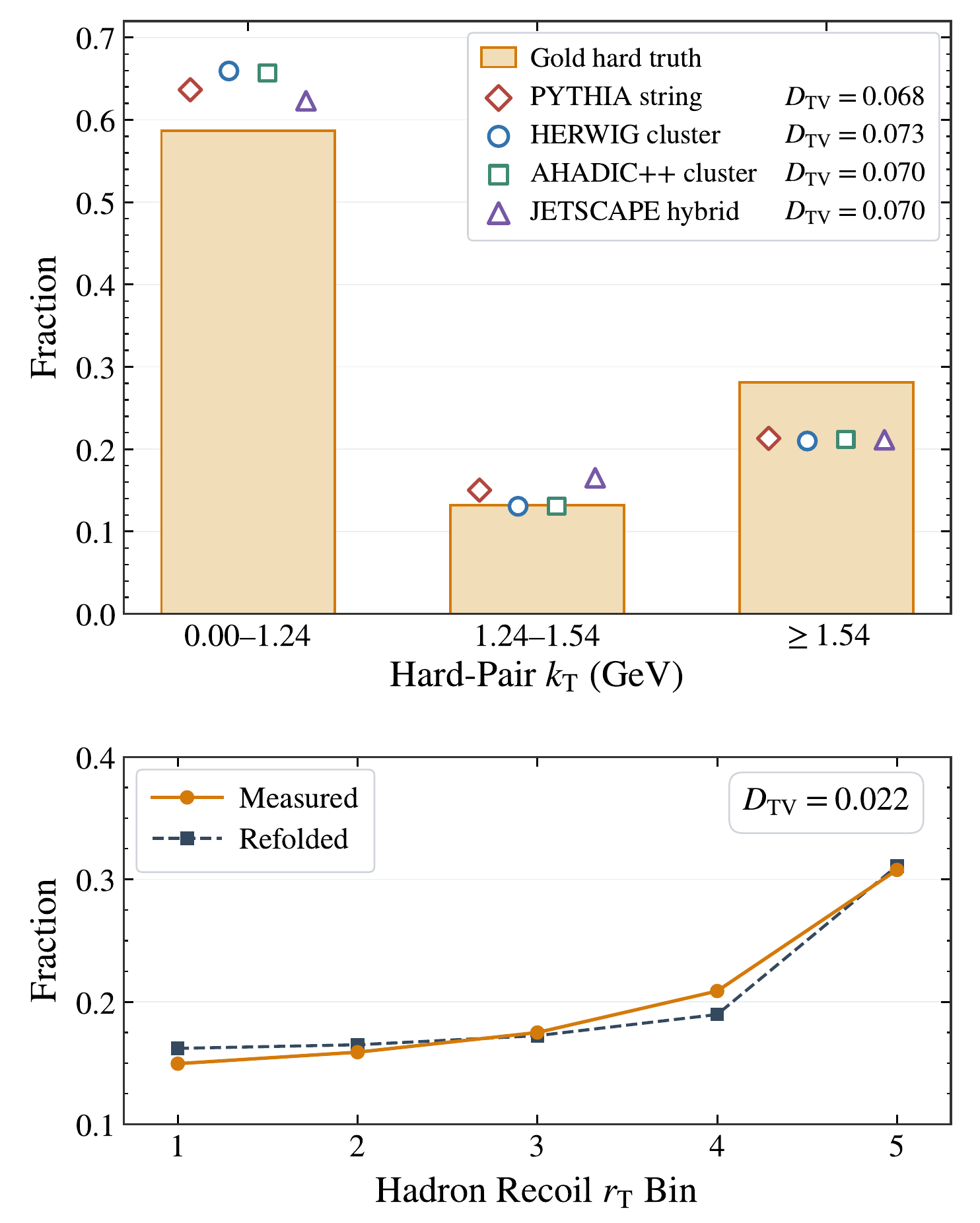}
  \caption{Reconstruction of the transverse-momentum imbalance of the hard $q\bar q$ pair in $e+\mathrm{Au}$ using $e+p$ response matrices. Upper panel: generated $k_T$ distribution and Bayesian reconstructions for four hadronization models. Lower panel: particle-level $r_T$ distributions from direct simulation and from propagating the reconstructed partonic distribution through $R^p$ for the PYTHIA string sample.}
  \label{fig:ibu-reconstruction}
\end{figure}

\bigskip

\noindent\textit{\textbf{Summary.}}
Revealing gluon saturation is one of the central goals of experiments at the EIC. A key challenge is to connect nonlinear small-$x$ parton dynamics to measurable quantities in hadronic final states after parton showers and hadronization redistribute the underlying transverse momentum. We address this challenge with a fully exclusive event-generator study and introduce global hadronic recoil as a new measurement strategy. In our simulations, the global recoil retains sensitivity to gluon saturation that is otherwise strongly diluted in leading-hadron and angular energy-flow observables. This sensitivity is evident at the hadron level in the persistent target dependence of $C_T$ throughout the displayed $x_B$ range. Bayesian unfolding of the global recoil using an $e+p$ response further reconstructs the coarsely binned $k_T$ distribution of the hard $q\bar q$ pair in $e+\mathrm{Au}$ for all four hadronization prescriptions considered. Global hadronic recoil thus connects nonlinear QCD dynamics to experimentally accessible multiparticle final-state observables, opening a new path to reveal gluon saturation after hadronization.

\noindent\textit{\textbf{Note added.}}---While this work was being finalized, Ref.~\cite{Duan:2026jrk} appeared, presenting a related study of gluon saturation in DIS.

\bigskip

\noindent\textit{Acknowledgments.}
The authors thank Shu-Yi Wei for helpful discussions. The work is supported by the National Natural Science Foundation of China under Grant Nos.~12175118 and 12321005 (J.~Z.), 12605155 (Y.~S.), 12575140 (W.~K.) and 12535010 (X.-N. W).
\clearpage
\onecolumngrid

\appendix
\setcounter{secnumdepth}{1}
\section*{Appendices}

To embed the nonlinear small-$x$ shower in eHIJING~\cite{Ke:2023xeo}, we
partition the hard-scattering phase space at $x_{\mathrm{sw}}=0.01$.  The
events analyzed here lie at $x_g<x_{\mathrm{sw}}$ and are sampled from the
off-shell $\gamma^*g^*\to q\bar q$ cross section convoluted with the evolved
gluon TMD, while events at larger $x_g$ are generated from collinear-factorized
hard cross sections and evolved with conventional \textsc{Pythia} ISR.  For
each small-$x$ event, the backward shower follows GLR evolution below
$x_{\mathrm{sw}}$~\cite{Shi:2022hee,Shi:2023ejp} and conventional DGLAP
evolution at larger $x$.  The hard partons and ISR emissions are then combined
with a momentum-balancing target remnant, organized into color-singlet chains,
and passed to final-state showering and hadronization.

Appendix~\ref{app:folded-glr} presents the folded GLR evolution, and
Appendix~\ref{app:hard-generation} gives the off-shell hard
process $\gamma^*g^*\to q\bar q$.  The backward initial-state radiation (ISR)
shower, its GLR--DGLAP matching, and the construction of fully exclusive
hadronic final states are described in
Appendices~\ref{app:backward-glr}--\ref{app:event-assembly}.
Appendix~\ref{app:ibu} details the iterative Bayesian unfolding and the
comparisons between reconstructed and directly generated distributions.

Key notation and symbols are summarized below.
\begin{center}
\begin{tabular}{ll}
\hline
Symbol & Physical meaning \\
\hline
$\mathcal{N}_A(\eta,k)$ & Gluon density (dimensionless, folded GLR) \\
$\eta=\ln(x_{\mathrm{sw}}/x_g)$ & Small-$x$ rapidity variable ($x_{\mathrm{sw}}=0.01$) \\
$C_{\mathrm{nl}}=5$ & Nonlinear GLR coefficient (fusion loss strength) \\
$S_{\perp,A}$ & Target transverse area ($\mathrm{GeV}^{-2}$) \\
$X=(x_g,k_T)$ & Joint generator-level coordinate for unfolding \\
$Y=(x_B,Q^2,r_T)$ & Joint particle-level observable coordinate \\
$r_T=|\sum_{-2<y_h<2}\boldsymbol{p}_{T,h}|$ & Global hadronic vector recoil entering $Y$ \\
$P_p(Y\mid X)$ & Response constructed from simulated $e+p$ events \\
$p_{T0}=2\,\mathrm{GeV}$ & Infrared regulator (GLR and DGLAP) \\
$x_{\mathrm{sw}}=0.01$ & GLR--DGLAP switch point \\
$D_{\mathrm{TV}}$ & Difference measure for normalized distributions \\
$\kappa$ & Response-matrix condition number \\
$\theta_h,\,y_h,\,\phi_h$ & Breit-frame hadron polar angle, rapidity, azimuth \\
\hline
\end{tabular}
\end{center}


\section{Folded GLR evolution}
\label{app:folded-glr}
\label{app:architecture}

The folded GLR evolution equation~\cite{Shi:2022hee,Shi:2023ejp} forms the
basis of our parton-shower algorithm.  We evolve the dimensionless
momentum-space gluon density $\mathcal{N}_A(\eta,\boldsymbol{k})$ in the
rapidity variable $\eta=\ln(x_{\mathrm{sw}}/x_g)$, with
$x_{\mathrm{sw}}=0.01$.  Assuming azimuthal symmetry, we write
$\mathcal{N}_A(\eta,\boldsymbol{k})=\mathcal{N}_A(\eta,k)$, where
$k=|\boldsymbol{k}|$.  The density is related to the target gluon TMD by
\begin{equation}
\mathcal{N}_A(\eta,\boldsymbol k)=\frac{2\alpha_s\pi^3}{N_c\, S_{\perp,A}}\,x_gG_A(\eta, \boldsymbol k),
\label{eq:supp-normalization}
\end{equation}
where $S_{\perp,A}$ is the target transverse area and $x_gG_A$ denotes the
gluon TMD.  With two-gluon fusion as the only nonlinear correction,
$\mathcal{N}_A(\eta,\boldsymbol{k})$ obeys the GLR evolution equation.
With running-coupling and kinematical-constraint effects included, this
equation reads~\cite{Shi:2022hee,Shi:2023ejp}
\begin{equation}
 \frac{\partial\mathcal{N}_A(\eta,\boldsymbol k)}{\partial\eta}
 =\bar\alpha_s(\boldsymbol k)
 \int 
 \frac{d^2\boldsymbol q}{\pi \boldsymbol q^2}
 \left[
   \mathcal{N}_A \left(\eta'=\eta+\ln\frac{\boldsymbol k^2}{\boldsymbol k^2+\boldsymbol q^2},\lvert\boldsymbol k+\boldsymbol q\rvert \right)
   -\Theta(\boldsymbol k-\boldsymbol q)\mathcal{N}_A(\eta,\boldsymbol k)
 \right]
 -C_{\mathrm{nl}}\bar\alpha_s(\boldsymbol k)\mathcal{N}_A^2(\eta,\boldsymbol k),
 \label{eq:supp-folded-glr}
\end{equation}
with $\bar\alpha_s(\boldsymbol k)= \alpha_s(\boldsymbol k)N_c/\pi$. The first term on the right-hand side describes real gluon emission. The kinematical constraint introduces a rapidity shift in the real-emission contribution, $\eta'=\eta+\ln\frac{\boldsymbol k^2}{\boldsymbol k^2+\boldsymbol q^2}$. The second term represents the virtual contribution, while the last term describes the nonlinear loss due to gluon recombination. The nonlinear coefficient $C_{\mathrm{nl}}$ parametrizes the strength of the gluon-fusion contribution~\cite{Motyka:2023pmt}.

For the shower construction, we write Eq.~\eqref{eq:supp-folded-glr} in gain--loss form,
\begin{equation}
\partial_\eta\mathcal{N}_A=\mathcal{R}_A-\Gamma_A\mathcal{N}_A,
\end{equation}
with
\begin{align}
 \mathcal{R}_A(\eta,\boldsymbol k)=\bar\alpha_s(\boldsymbol k)
 \int_{q_{\min}}^{q_{\max}(\eta,\boldsymbol k)}
 \frac{d^2\boldsymbol q}{\pi q^2}
 \mathcal{N}_A(\eta',\lvert\boldsymbol k+\boldsymbol q\rvert),\quad 
 \Gamma_A(\eta,\boldsymbol k)=\bar\alpha_s(\boldsymbol k)\left[
 \int_{q_{\min}}
 \frac{d^2\boldsymbol q}{\pi \boldsymbol q^2}\Theta(\boldsymbol k-\boldsymbol q)
 +C_{\mathrm{nl}}\mathcal{N}_A(\eta,\boldsymbol k)\right].
 \label{eq:supp-real-loss}
\end{align}
Here $\mathcal{R}_A$ is the real-emission gain evaluated at the shifted
rapidity, while $\Gamma_A$ combines the virtual and nonlinear fusion losses.
The infrared cutoff $q_{\min}$ and ultraviolet cutoff $k_{\max}$ are introduced
for the numerical solution; the result is insensitive to their values.  The
real-emission phase space is bounded by
$q_{\max}(\eta,\boldsymbol{k})=\min\left[k\sqrt{x_{\mathrm{sw}}/x_g-1},k_{\max}\right]$;
the first bound follows from $\eta'>0$ in Eq.~\eqref{eq:supp-folded-glr}.
Exponentiating the loss rate gives the formal survival factor for no resolvable
branching or fusion between $\eta_1$ and $\eta_2$,
\begin{equation}
\Delta(\eta_2,\eta_1;k)=\exp\left[-\int_{\eta_1}^{\eta_2}d\xi\,\Gamma_A(\xi,k) \right],
 \label{eq:supp-Detla}
\end{equation}
and hence the folded integral equation
\begin{equation}
 \mathcal{N}_A(\eta_2,k)=
\Delta(\eta_2,\eta_1;k)\mathcal{N}_A(\eta_1,k)
 +\int_{\eta_1}^{\eta_2}d\xi\,
\Delta(\eta_2,\xi;k)\mathcal{R}_A(\xi,k).
 \label{eq:supp-folded-integral}
\end{equation}
The first term propagates the density at $\eta_1$ without a resolvable
branching or fusion, while the second adds real emissions generated at each
intermediate rapidity $\xi$ and subsequently evolved to $\eta_2$.  We solve
Eq.~\eqref{eq:supp-folded-integral} forward from $\eta_0$ to $\eta_{\max}$.
This solution provides the density and loss integral used to construct the
regulated survival probability for the backward shower.

We use the parent-momentum prescription for the running coupling,
\begin{equation}
\alpha_s(\mu)=\frac{4\pi}
{\beta_0\ln\left(\frac{C_0^2\mu^2+m_0^2}{\Lambda^2}\right)},
\end{equation}
where $\beta_0=11-\frac{2}{3}N_f$, $C_0=2.5$,
$\Lambda^2=0.0578\,\mathrm{GeV}^2$, and
$m_0^2=0.425\,\mathrm{GeV}^2$.  We set $C_{\mathrm{nl}}=5$.  For the numerical
implementation, we introduce an infrared transverse-momentum scale $p_{T0}$
to ensure smooth matching to \textsc{Pythia} at $x=0.01$ and
$Q^2=25\,\mathrm{GeV}^2$.  We make the replacement
$\boldsymbol{k}^2\rightarrow \boldsymbol k^2+p_{T0}^2$, with
$p_{T0}=2\,\mathrm{GeV}$, in both the running coupling and the denominator
entering the evolution kernel.  With this prescription, the factor defined in
Eq.~\eqref{eq:supp-Detla} can be expressed analytically as
\begin{equation}
\Delta(\eta_2,\eta_1;\boldsymbol k)  =\exp\left\{-\bar\alpha_s\left(\sqrt{\boldsymbol k^2+p_{T0}^2}\right)
 \left[C_{\mathrm{nl}}\int_{\eta_1}^{\eta_2}d\eta\,\mathcal{N}_A(\eta,\boldsymbol k)
 +(\eta_2-\eta_1)\ln \frac{\boldsymbol k^2+p_{T0}^2}{\boldsymbol q_{\min}^2+p_{T0}^2}\right]\right\}.
 \label{eq:supp-survival} 
\end{equation}
At the starting rapidity $\eta_0=0$, both targets are initialized with the Gaussian profile
\begin{equation}
\mathcal{N}_{A/p}(0,k)=\mathcal{N}_{0,A/p}\exp\left(-\frac{k^2}{Q_{s0}^2}\right).    
\end{equation}
For the proton, we choose $Q_{s0,p}^2=0.64\,\mathrm{GeV}^2$ and
$S_{\perp,p}=102\,\mathrm{GeV}^{-2}$; the former is consistent with the
nonperturbative transverse-momentum broadening implemented in \textsc{Pythia}.
To fix $\mathcal{N}_{0,p}$, we match the gluon TMD implied by
Eq.~\eqref{eq:supp-normalization} to the collinear gluon PDF at $x=0.01$ and
$\mu^2=25\,\mathrm{GeV}^2$:
\begin{equation*}
\int d^2\boldsymbol{k}\, x_g G_A(0,\boldsymbol{k})
= \frac{N_cS_{\perp,p}}{2\alpha_s\pi^3}\int d^2\boldsymbol{k}\,
\mathcal{N}_p(0,\boldsymbol{k})=xg(x,\mu^2).
\end{equation*}
This gives $\mathcal{N}_{0,p}=0.142$.  For Au, we take
$Q_{s0,A}^2=4\,\mathrm{GeV}^2$, $\mathcal{N}_{0,A}=0.148$, and
$S_{\perp,A}=3357\,\mathrm{GeV}^{-2}$, with the normalization fixed by
$xG_{\mathrm{Au}}/xG_p=197$ at $x=0.01$ and $Q^2=25\,\mathrm{GeV}^2$.


\section{Quark-pair production cross section in DIS}
\label{app:hard-generation}

The transverse-momentum imbalance of a quark-antiquark pair in DIS directly
probes the $k_T$ dependence of the gluon TMD. We consider the process
\begin{equation}
l+A\rightarrow l'+q+\bar q+X.
\end{equation}
Let $P$, $l$, and $q=l-l'$ denote, respectively, the incoming hadron momentum
per nucleon, the incoming lepton momentum, and the exchanged virtual-photon
momentum.  The DIS variables are $Q^2=-q^2$, $x_B=Q^2/(2P\cdot q)$, and
$y=(P\cdot q)/(P\cdot l)$.  The quark-antiquark pair is produced through the
interaction of the exchanged virtual photon $\gamma^*$ with the target.  The
pair carries the transverse momentum of the off-shell gluon,
$\boldsymbol{k}_T=\boldsymbol{k}_{1T}+\boldsymbol{k}_{2T}$.  Here
$\boldsymbol{k}_{1T}$ and $\boldsymbol{k}_{2T}$ are the transverse momenta of
the quark and antiquark, respectively, while $z_q$ is the photon light-cone
momentum fraction carried by the quark.
In the one-gluon-exchange approximation, the corresponding cross section is given by
\begin{equation}
\frac{d\sigma}
{dQ^2\,dx_B\,dz_q\,d^2\boldsymbol{k}_{1T}\,d^2\boldsymbol{k}_{2T}}
=
\frac{\alpha_{\mathrm{em}}^2 N_c S_{\perp,A}}
{2\pi^4 x_B Q^2}
\sum_f e_f^2\,
\mathcal{H}_f
(Q^2,x_B,z_q,\boldsymbol{k}_{1T},\boldsymbol{k}_{2T})
\frac{\mathcal{N}_A(\eta,\boldsymbol{k}_T)}
{\boldsymbol{k}_T^2}.
\label{eq:supp-hard-weight}
\end{equation}
The hard factor is given by
\begin{align}
\mathcal{H}_f
(Q^2,x_B,z_q,\boldsymbol{k}_{1T},\boldsymbol{k}_{2T})
={}&
\left(1-y+\frac{y^2}{2}\right)
\Bigg\{
[z_q^2+(1-z_q)^2]
\left(
\frac{\boldsymbol{k}_{1T}}
{\boldsymbol k_{1T}^2+z_q(1-z_q)Q^2+m_f^2}
+
\frac{\boldsymbol{k}_{2T}}
{\boldsymbol k_{2T}^2+z_q(1-z_q)Q^2+m_f^2}
\right)^2
\nonumber\\
&\qquad
+m_f^2
\left(
\frac{1}
{\boldsymbol k_{1T}^2+z_q(1-z_q)Q^2+m_f^2}
-
\frac{1}
{\boldsymbol k_{2T}^2+z_q(1-z_q)Q^2+m_f^2}
\right)^2
\Bigg\}
\nonumber\\
&+
4(1-y)z_q^2(1-z_q)^2Q^2
\left(
\frac{1}
{\boldsymbol k_{1T}^2+z_q(1-z_q)Q^2+m_f^2}
-
\frac{1}
{\boldsymbol k_{2T}^2+z_q(1-z_q)Q^2+m_f^2}
\right)^2 .
\label{eq:supp-hard-factor}
\end{align}
Here $m_f$ is the quark mass, and $\eta=\ln(x_{\rm sw}/x_g)$.  The gluon
momentum fraction $x_g$ and the quark-pair invariant mass $M_{q\bar q}$ are
defined as 
\begin{equation}
x_g=x_B\left[
1+\frac{m_f^2+\boldsymbol k_{1T}^2}{z_qQ^2}
+\frac{m_f^2+\boldsymbol k_{2T}^2}{(1-z_q)Q^2}
\right],
\qquad
M_{q\bar q}^2=
\frac{m_f^2+\boldsymbol k_{1T}^2}{z_q}
+\frac{m_f^2+\boldsymbol k_{2T}^2}{1-z_q}
-\left|\boldsymbol{k}_{1T}+\boldsymbol{k}_{2T}\right|^2 .
\label{eq:supp-hard-xg}
\end{equation}
In the dilute limit, our result agrees with that of
Ref.~\cite{Dominguez:2011br}.  A key ingredient is the factor
$\mathcal{N}_A(\eta,\boldsymbol{k}_T)/\boldsymbol{k}_T^2$:
$\mathcal{N}_A$ is the evolved gluon density evaluated at the sampled
longitudinal momentum fraction and transverse momentum, while
$1/\boldsymbol{k}_T^2$ is the gluon propagator factor.  In the
$\boldsymbol{k}_T\to0$ limit, the vanishing hard factor compensates the
apparent gluon-propagator singularity, leaving the cross section finite.

For Monte Carlo event generation, we sample the final-state phase space according
to the differential cross section in Eq.~\eqref{eq:supp-hard-weight}.  We choose
the factorization scale $\mu_{\rm fa}=\sqrt{p_{T,\mathrm{fac}}^2+m_f^2}$, where
$\boldsymbol p_{T,\mathrm{fac}}=(1-z_q)\boldsymbol k_{1T}-z_q\boldsymbol k_{2T}$
is the relative momentum of the pair.  The numerical weight retains $m_f$ in
the kinematics and propagator denominators but omits the separate term
proportional to $m_f^2$ in Eq.~\eqref{eq:supp-hard-factor}.  We denote the sampled
variables collectively by
 \begin{equation}
     t=(x_B,Q^2,z_q,\boldsymbol k_{1T},\boldsymbol k_{2T}),
 \end{equation}
 from which the full final-state kinematics can be reconstructed.

For the phenomenological study of light-quark pair production, we consider
$18\,\mathrm{GeV}$ electrons colliding with hadron beams of
$110\,\mathrm{GeV}$ per nucleon and impose the kinematic cuts $0<y<1$,
$0<x_g<0.01$, and $Q^2\ge2\,\mathrm{GeV}^2$.

The gluon momentum fraction $x_g$ is reconstructed from
Eq.~\eqref{eq:supp-hard-xg} and determines the evolution regime. For
$x_g<0.01$, the gluon density is described by the GLR evolution equation in
Eq.~\eqref{eq:supp-folded-integral}. We employ the backward-evolution algorithm
developed in Refs.~\cite{Shi:2022hee,Shi:2023ejp} to generate the initial-state
gluon radiation associated with the small-$x$ evolution, as described in the
next section. Once backward evolution reaches momentum fractions above
$x_{\mathrm{sw}}=0.01$, where the small-$x$ evolution is no longer applicable,
we switch to a conventional collinear parton shower, implemented with
\textsc{Pythia}, to account for initial-state radiation in the large-$x$
region. Further details of the matching between the two evolution regimes are
given below.

\section{Backward GLR shower and small-$x$ ISR}
\label{app:backward-glr}

The backward GLR shower reconstructs the small-$x$ initial-state gluon
radiation leading to the off-shell gluon selected by the hard scattering.  The
shower starts from the current $t$-channel gluon state,
$x_c=x_{\mathrm{sw}}e^{-\eta_c}$ and $(k_c,\phi_c)=(k_T,\phi_g)$.  Here
$x_{\mathrm{sw}}=0.01$, and the subscript $c$ denotes the current state.
The starting rapidity is restricted to $0\le\eta_c\le6$.  The gluon
density, its rapidity integral, and the no-branching probability are the
target-specific quantities defined in Sec.~\ref{app:folded-glr}.

The next trial branching rapidity is generated from the no-branching probability
$B_A(\eta,\eta_c;k_c)$, which is defined as~\cite{Shi:2022hee,Shi:2023ejp}
\begin{equation}
B_A(\eta,\eta_c;k)=\Delta(\eta_c,\eta;k)
\frac{\mathcal{N}_A(\eta,k)}{\mathcal{N}_A(\eta_c,k)}.    
\end{equation}
At fixed $(\eta_c,k_c)$, the algorithm draws a uniform random number $r$ and
scans toward smaller $\eta$ on a grid with spacing $\Delta\eta=0.03$ to find the
first interval in which $r>B_A$.  The trial branching rapidity is then obtained by interpolation
within that interval.
If no crossing occurs before $\eta=0$, the small-$x$ stage ends without a
further emission.  The boundary extrapolation that initializes the subsequent
evolution is not counted as a GLR emission.

A trial branching at rapidity $\eta_v$ with transverse momentum
$\boldsymbol l=(l,\phi_l)$ defines the parent $t$-channel gluon state
$\boldsymbol k_p=\boldsymbol k_c-\boldsymbol l$, with
$k_p=|\boldsymbol k_p|$, $\eta_p=\eta_v+\ln[k_c^2/(k_c^2+l^2)]$, and
$x_p=x_{\mathrm{sw}}e^{-\eta_p}$.
The real-emission veto uses the density at
$\eta_{\mathrm{kin}}=\max(0,\eta_p)$ and $k_p$.
The trial transverse momentum is sampled from a distribution defined by the
overestimate
$\mathcal{P}_{\mathrm{env}}=\mathcal{E}(k_c,\eta_v)/[|k_c^2-l^2|+\delta_{\mathrm{prop}}^2]$,
with $\delta_{\mathrm{prop}}^2=1.5\,\mathrm{GeV}^2$, where $\mathcal{E}$ is a
state-local envelope.
The acceptance probability is
\begin{equation}
 P_{\mathrm{acc}}=
 \frac{\mathcal{N}_A(\eta_{\mathrm{kin}},k_p)
       \,l^2/(l^2+p_{T0}^2)}
       {\mathcal{P}_{\mathrm{env}}},
 \qquad p_{T0}=2\,\mathrm{GeV}.
 \label{eq:supp-bk-acceptance}
\end{equation}
The envelope is chosen so that $\mathcal{P}_{\mathrm{env}}$ bounds the target
weight for every trial $(l,\phi_l)$, equivalently
\begin{equation*}
 \mathcal{E}(k_c,\eta_v)\ge
 \mathcal{N}_A(\eta_{\mathrm{kin}},k_p)
 \frac{l^2}{l^2+p_{T0}^2}
 \left(|k_c^2-l^2|+\delta_{\mathrm{prop}}^2\right).
\end{equation*}
Numerically, this bound is constructed from cellwise maxima of the tabulated
density, with an additional bound to account for clamping at the high-$k_p$
edge.  This
construction ensures $P_{\mathrm{acc}}\le1$ throughout phase space;
any violation of this bound terminates event generation.

An accepted branching must satisfy $0<x_c<x_p<1$, with the emitted gluon
carrying $x_e=x_p-x_c$.  In per-nucleon light-cone coordinates,
\begin{equation}
 p_e^+=\frac{x_e\sqrt{s_{NN}}}{\sqrt{2}},
 \qquad
 p_e^-=\frac{l^2}{2p_e^+},
 \qquad
 (p_{T,e},\phi_e)=(l,\phi_l),
 \label{eq:supp-emitted-gluon}
\end{equation}
so that $p_e^2=0$.
For each accepted branching, the color flow is updated by inserting the emitted
gluon on a randomly chosen side of the current color line; the gluon and the new
spacelike parent receive complementary color pairs.  The emitted gluon remains in the small-$x$
initial-state cascade, while the parent, with virtuality $-k_p^2$, becomes the
state for the next backward step.

\section{GLR--DGLAP matching and larger-$x$ evolution}
\label{app:dglap-handoff}

The small-$x$ evolution is initialized at $x_{\mathrm{sw}}=0.01$.  For
inclusive observables, dynamics at larger $x$ can be encoded in this initial
condition.  Fully exclusive events, however, require an explicit initial-state
radiation history also above $x_{\mathrm{sw}}$.  We therefore continue the
initial-state shower with gluon-only DGLAP evolution.

The matching at $x_{\mathrm{sw}}=0.01$ partitions the backward initial-state
shower without overlap.  Starting at
$\eta_c=\ln(x_{\mathrm{sw}}/x_g^{\mathrm{hard}})$, the GLR shower generates all
branchings whose current $t$-channel gluon satisfies $x_c<x_{\mathrm{sw}}$.  The first
accepted branching with $x_c<x_{\mathrm{sw}}\le x_p$ crosses the matching
boundary and is retained as the final GLR emission.  The GLR description,
including the no-branching probability and nonlinear loss term, ends at this
boundary; backward evolution then continues toward larger $x$ with a
gluon-only DGLAP shower.  If the matching boundary is reached without such a
crossing,
or if $k_c\le0.2\,\mathrm{GeV}$ or the parent momentum fraction is invalid, GLR
evolution ends without an additional emission.

The DGLAP shower begins from one of two matching configurations.  For a
crossing emission, the parent state $(x_p,k_p,\phi_p,\mathcal{C}_p)$ provides
the starting configuration at the scale $Q^2$ and sets the lower
momentum-fraction bound to $x_{\min}=x_p$.  If GLR evolution instead reaches $\eta=0$ without an
emission, only the last interval is extrapolated to define
$x_{\min}\ge x_{\mathrm{sw}}$.  This extrapolation creates no physical emission,
and DGLAP starts from $x_g^{\mathrm{hard}}$ at the scale $Q^2$.

In either case, the first DGLAP trial branching must satisfy
$x_p\ge x_{\min}\ge x_{\mathrm{sw}}$.
A trial branching below this boundary is rejected to prevent overlap with the
GLR phase space, and the shower terminates at the matching boundary.

Above the matching boundary, the gluon-only DGLAP shower uses the unregularized
real-emission splitting kernel
$P_{gg}(z)=2C_A[z/(1-z)+(1-z)/z+z(1-z)]$, with $C_A=3$.
The veto algorithm uses the overestimate
$P_{gg}^{\mathrm{over}}(z)=2C_A[1/(1-z)+1/z]$,
with the proton gluon-PDF ratio
$xg(x_p,\mu^2_{\mathrm{trial}})/xg(x_c,\mu^2_{\mathrm{trial}})$ entering the
acceptance probability.
With one-loop running fixed by $\alpha_s(m_Z^2)=0.1365$ and no CMW rescaling,
the strong coupling is evaluated at $\mu^2_{\mathrm{trial}}+p_{T0}^2$.

Infrared behavior is regulated through the shower evolution measure
$dP \sim d\mu^2/(\mu^2+p_{T0}^2)$ with $p_{T0}=2\,\mathrm{GeV}$ and the cutoff
$\mu^2_{\min}=(0.2\,\mathrm{GeV})^2$.  Using the same $p_{T0}$ in the GLR
virtual term provides a common infrared prescription across the matching
boundary.

Successive DGLAP branchings $i$ and $i+1$ are subject to the rapidity-ordering
constraint
\begin{equation}
 \mu^2_{i+1} \le \left[\frac{1-z_{i+1}}{z_{i+1}(1-z_i)}\right]^2\mu^2_i.
 \label{eq:supp-dglap-ordering}
\end{equation}
The first DGLAP trial branching is ordered with respect to the
photon--gluon-fusion configuration $(z_h,\mu_h^2)$ supplied by the hard event,
even when GLR produces no emission.  The shower terminates when $\mu^2$ falls
below the infrared cutoff or the parent momentum fraction reaches its kinematic
limit of unity.  The matching constraint above applies only to the first trial
branching.

\section{Event assembly and hadronization}
\label{app:event-assembly}

The hadronization input contains the hard $q\bar q$ pair and the real
initial-state emissions generated during backward evolution.  Unlike in a
standard recoil-based shower, subsequent branchings do not redistribute
transverse momentum among previously generated partons.
The terminal spacelike gluon marks the endpoint of backward evolution and is
not passed to hadronization.  We label each emission by its GLR or DGLAP origin
and assemble the partonic state as
\begin{equation}
 \{p_q^{\mathrm{hard}},p_{\bar q}^{\mathrm{hard}}\}
 \cup\{p_j^{\mathrm{GLR}}\}
 \cup\{p_j^{\mathrm{DGLAP}}\}.
 \label{eq:supp-parton-state}
\end{equation}
We write the ISR momenta in light-cone components,
\begin{equation}
 p^\pm=\frac{E\pm p_z}{\sqrt2},
 \qquad
 E=\frac{p^++p^-}{\sqrt2},
 \qquad
 p_z=\frac{p^+-p^-}{\sqrt2},
 \label{eq:supp-lightcone}
\end{equation}
with $p_x=p_T\cos\phi$ and $p_y=p_T\sin\phi$ completing each ISR four-vector.
We instead reconstruct the hard four-vectors in an orthonormal tetrad built
from $q+x_gP$.  Its timelike axis lies along $q+x_gP$, its longitudinal axis is
fixed by the component of $x_gP$ orthogonal to that direction, and the
lepton--hadron scattering plane fixes the remaining axes.  In this basis, an
on-shell momentum with mass $m$, transverse momentum $\boldsymbol k_T$, and
positive minus component $k^-$ satisfies
\begin{equation}
 k^+=\frac{m^2+k_T^2}{2k^-},
 \label{eq:supp-onshell-tetrad}
\end{equation}
with Cartesian components obtained from Eq.~\eqref{eq:supp-lightcone}.
The quark and antiquark have $k_1^-=z_h q^-$ and
$k_2^-=(1-z_h)q^-$ in this basis before being transformed back to the
laboratory frame.

To leave the target-specific $q\bar q$ four-vectors unchanged, we initialize
backward evolution from a separate auxiliary two-parton configuration whose
local $x$ axis is aligned with the relative momentum $p_{\mathrm{rel}}$ of the pair.  In
this frame, the pair transverse momentum is
\begin{equation}
 \boldsymbol q_T=k_T(\cos\phi_{pq},\sin\phi_{pq}),
 \label{eq:supp-dijet-qT}
\end{equation}
and the auxiliary transverse momenta are
\begin{equation}
 \boldsymbol k_{1T}=(p_{\mathrm{rel}},0)+\tfrac12\boldsymbol q_T,
 \qquad
 \boldsymbol k_{2T}=(-p_{\mathrm{rel}},0)+\tfrac12\boldsymbol q_T,
 \label{eq:supp-dijet-kT}
\end{equation}
so that $\boldsymbol k_{1T}+\boldsymbol k_{2T}=\boldsymbol q_T$.  Together
with the recorded rapidities, the transverse masses
$m_{Ti}=\sqrt{m_f^2+k_{iT}^2}$ determine the light-cone components
\begin{equation}
 k_i^+=\frac{m_{Ti}e^{y_i}}{\sqrt2},
 \qquad
 k_i^-=\frac{m_{Ti}e^{-y_i}}{\sqrt2}.
 \label{eq:supp-dijet-lightcone}
\end{equation}
The hard $x_g$ and the positive light-cone fractions of the real initial-state
emissions determine the reconstructed incoming momentum fraction
$x_{\mathrm{par}}$.  Exact momentum conservation then fixes the total target
remnant momentum:
\begin{equation}
 p_R^+=(1-x_{\mathrm{par}})\sqrt2 E_P,
 \qquad
 \boldsymbol t_T=\boldsymbol q_T-
 \sum_{i\in\{q,\bar q,\mathrm{ISR}\}}\boldsymbol p_{T,i},
 \label{eq:supp-remnant-total}
\end{equation}
where $\boldsymbol t_T$ ensures transverse-momentum conservation with the
exchanged boson.  We represent the remnant by an on-shell quark and diquark.
For a sampled momentum partition $z_R$, their components are
\begin{align}
 p_q^+&=(1-z_R)p_R^+,
 &\boldsymbol p_{T,q}&=(1-z_R)\boldsymbol t_T,
 \nonumber\\
 p_d^+&=z_R p_R^+,
 &\boldsymbol p_{T,d}&=z_R\boldsymbol t_T,
 \label{eq:supp-remnant-split}
\end{align}
and their minus components follow from
\begin{equation}
 p_i^-=\frac{m_i^2+p_{T,i}^2}{2p_i^+}.
 \label{eq:supp-remnant-minus}
\end{equation}
We assign no additional intrinsic transverse momentum to the remnant
endpoints and sample $z_R$ from a Beta$(1/2,1)$ distribution.  For the nominal
constituent masses and spin assignments, the endpoints are $d+(uu)_1$ with
probability $1/3$ and $u+(ud)$ otherwise.

For hadronization, we arrange the partons into two color-singlet chains,
\begin{equation}
 q_{\mathrm{hard}}\longrightarrow
 \mathrm{ISR}_1\longrightarrow\cdots\longrightarrow
 \mathrm{ISR}_n\longrightarrow d_{\mathrm{rem}},
 \qquad
 q_{\mathrm{rem}}\longrightarrow\bar q_{\mathrm{hard}}.
 \label{eq:supp-color-chains}
\end{equation}
The initial-state emissions retain their GLR-then-DGLAP order along the first
chain, while the color assignment leaves all sampled momenta unchanged.  We
retain the two-chain state for showering only if
\begin{equation}
 \left|\frac{\sum_{i\in\mathrm{chain}}E_i}{E_P+q^0}-1\right|\le0.20,
 \label{eq:supp-energy-check}
\end{equation}
where $\mathrm{chain}$ denotes the union of the two color-singlet chains in
Eq.~\eqref{eq:supp-color-chains}.
To satisfy this bound, we redraw only $z_R$, keeping the chosen remnant flavors
fixed, and discard the event after 100 unsuccessful attempts.

Final-state radiation (FSR) acts on the accepted colored partonic state while
the hard process and initial-state shower remain fixed.  We do not include
beam intrinsic transverse momentum.  Only the hard quark and antiquark initiate
time-like radiation; the ISR partons and remnant endpoints start at the
negligible scale $Q_{\min}$.  The local scales are
\begin{equation}
 Q_{i,\mathrm{FSR}}=
 \begin{cases}
 \max\!\left(\sqrt{p_{T,\mathrm{fac}}^2+m_f^2},p_{\mathrm{rel}},Q_{\min}\right),
   &i=q_{\mathrm{hard}},\bar q_{\mathrm{hard}},\\
 Q_{\min},&i=\mathrm{ISR}\ \text{or remnant},
 \end{cases}
 \qquad Q_{\min}=10^{-6}\,\mathrm{GeV}.
 \label{eq:supp-fsr-scales}
\end{equation}
The largest of these values sets the upper scale of the time-like shower
before hadronization.

This construction maintains a one-to-one correspondence between the
perturbative and hadronized records.  It preserves the hard-event kinematics
and leaves unchanged the numbers and ordering of hard events, shower starting
states, and emission histories.  The tetrad used to reconstruct the hard momenta
remains orthonormal, and every retained event satisfies the on-shell conditions,
the DIS identities $q^2=-Q^2$ and $2P\cdot q=Q^2/x_B$, and four-momentum
conservation within $10^{-5}$.


\section{Iterative Bayesian unfolding}
\label{app:ibu}

We use iterative Bayesian unfolding (IBU) to reconstruct the coarsely binned $k_T$ distribution of the hard $q\bar q$ pair from fully exclusive $e+p$ and $e+\mathrm{Au}$ final states. The response is determined entirely from
simulated $e+p$ events.

Particle-level events are binned in $Y=(x_B,Q^2,r_T)$, with global recoil
\begin{equation}
 r_T = \left|\sum_{h:\,|y_h|<2}\boldsymbol p_{T,h}\right|.
 \label{eq:supp-recoil-def}
\end{equation}
Generator-level events are binned in
\begin{equation}
 X=(x_g,k_T),\qquad
 k_T = |\boldsymbol k_{1T} + \boldsymbol k_{2T}|,
\label{eq:supp-hard-pair-imbalance}
\end{equation}
where $\boldsymbol k_{iT}$ denote the transverse momenta of the quark and
antiquark produced in the off-shell process $\gamma^*g^*\to q\bar q$
(Sec.~\ref{app:hard-generation}).
The scattered-lepton four-momentum determines $x_B$ and $Q^2$, whereas $r_T$
includes all stable hadrons in $-2<y_h<2$.  Within the selected $x_B$ range,
zero-recoil events are retained, and no cuts are imposed on hadron charge or
species.

Let $\alpha$ label a joint $(x_g,k_T)$ generator-level cell and $m$ a joint
$(x_B,Q^2,r_T)$ particle-level cell.  For simulated $e+p$ counts
$N_{m\alpha}$, the conditional response
$R^p_{m\alpha}=P_p(Y\in M_m\mid X\in B_\alpha)$ is
\begin{equation}
 R^p_{m\alpha}=\frac{N_{m\alpha}}{N_\alpha^{\mathrm{gen}}}.
\label{eq:supp-response-matrix}
\end{equation}
Because no detector simulation is performed and no detector-level selection is
applied, every generated event enters a particle-level bin.  The response therefore has unit efficiency and
each generator-level column sums to one.  Events without stable hadrons in the
recoil rapidity interval populate $r_T=0$; detector-level selection would
instead introduce an efficiency that depends on $(x_g,k_T)$.

Applying the response constructed from $e+p$ events to $e+\mathrm{Au}$ allows us
to examine the explicit approximation
$P_{\mathrm{Au}}(Y\mid X)\simeq P_p(Y\mid X)$ within the generator setup.  The
hard matrix element is common to the two targets at fixed $X=(x_g,k_T)$, while
the target-dependent backward shower can modify the conditional particle-level
distribution.  The joint response therefore includes the ISR-relevant $x_g$
dependence.  Its applicability to $e+\mathrm{Au}$ is determined by comparison
with the directly generated distribution.

For the target-dependent samples specified in Sec.~\ref{app:architecture}, 75\%
of the simulated $e+p$ events determine the response, initial prior, and all bin
edges; the remaining 25\% provide an independent $e+p$ sample.  Particle-level
counts from simulated $e+\mathrm{Au}$ events provide the IBU input.  The
resulting $40\times6$ response combines two $x_g$ bins and three $k_T$
intervals into six generator-level cells, and four $x_B$, two $Q^2$, and five
$r_T$ bins into 40 particle-level cells.  As listed in
Table~\ref{tab:supp-ibu-edges}, the three $k_T$ intervals are formed by grouping
sixteen fine quantile bins in a $14/1/1$ pattern, with 87.5\% of the events used
to construct the response in the first bin and 6.25\% in each tail bin.

\begin{table}[t]
 \caption{Nominal binning of the proton response
 $R^p_{m\alpha}=P_p(Y\in M_m\mid X\in B_\alpha)$.  The generator-level and
 particle-level spaces contain $6$ and $40$ cells, respectively.  All internal
 boundaries are fixed from the simulated $e+p$ events used to construct the
 response.  Momentum and virtuality edges are in GeV and $\mathrm{GeV}^2$;
 $\infty$ denotes the overflow boundary.  Values are rounded to the shown
 precision.}
 \label{tab:supp-ibu-edges}
 \begin{ruledtabular}
  \begin{tabular}{lc}
   Coordinate and response binning & Bin edges \\
   \hline
   \multicolumn{2}{c}{Generator level: $X=(x_g,k_T)$, $2\times3=6$ cells} \\
   $x_g$ (median split) & $0,\ 0.00424,\ \infty$ \\
   $k_T$ ($16$ quantiles grouped $14/1/1$)
     & $0,\ 1.24,\ 1.54,\ \infty$ \\
   \hline
   \multicolumn{2}{c}{Particle level: $Y=(x_B,Q^2,r_T)$, $4\times2\times5=40$ cells} \\
   $x_B$ (quartiles)
     & $0,\ 8.15\!\times\!10^{-4},\ 1.45\!\times\!10^{-3},\ 2.57\!\times\!10^{-3},\ \infty$ \\
   $Q^2$ (median split) & $0,\ 3.40,\ \infty$ \\
   $r_T$ (quintiles)
     & $0,\ 0.389,\ 0.607,\ 0.846,\ 1.21,\ \infty$ \\
  \end{tabular}
 \end{ruledtabular}
\end{table}

D'Agostini's iterative Bayesian method reconstructs the joint generator-level
distribution.  Let $\pi_\alpha^{(n)}$ be the normalized prior at iteration $n$
and $d_m$ the simulated particle-level event counts.  The initial prior is the
generator-level joint distribution of the same $e+p$ sample used to construct
the response.  The Bayesian inverse and update are
\begin{equation}
 B_{\alpha m}^{(n)}=
 \frac{R^p_{m\alpha}\pi_\alpha^{(n)}}
      {\sum_\beta R^p_{m\beta}\pi_\beta^{(n)}},
\label{eq:supp-bayes-inversion}
\end{equation}
\begin{equation}
 \widehat N_\alpha^{(n+1)}=\sum_m d_m B_{\alpha m}^{(n)},
 \qquad
 \pi_\alpha^{(n+1)}=
 \frac{\widehat N_\alpha^{(n+1)}}{\sum_\beta\widehat N_\beta^{(n+1)}},
\label{eq:supp-ibu-update}
\end{equation}
where the efficiency factor is unity for the all-event particle-level selection
used here.  The nominal analysis uses 20 iterations and no additional
regularization.  The final pair-imbalance distribution, integrated over $x_g$,
is
\begin{equation}
 \widehat N(k_T^b)=\sum_a \widehat N(x_g^a,k_T^b).
 \label{eq:supp-kt-distribution}
\end{equation}

For the independent 25\% $e+p$ sample, we quantify the agreement between the
reconstructed and generator-level joint distributions with
\begin{equation}
 D_{\mathrm{TV}}(p,q) = \frac12\sum_i|p_i-q_i|.
 \label{eq:supp-tv-distance}
\end{equation}
This independent proton sample yields $D_{\mathrm{TV}}^{e+p}=0.0078$ for the
joint distribution and $0.0036$ for the $k_T$ distribution integrated over
$x_g$.  Applying the same response to the simulated particle-level
$e+\mathrm{Au}$ sample gives
$D_{\mathrm{TV}}^{\mathrm{Au}}=0.068$ for both the joint distribution and
the $k_T$ distribution integrated over $x_g$; the $x_g$ distribution integrated
over $k_T$ gives $0.0098$.

Applying the $e+p$ response to the reconstructed $e+\mathrm{Au}$
generator-level distribution gives the corresponding particle-level
$(x_B,Q^2,r_T)$ bin vector
\begin{equation}
 d_m^{\mathrm{pred}} = \sum_\alpha R^p_{m\alpha}
 \widehat N_\alpha^{(20)}.
\label{eq:supp-particle-prediction}
\end{equation}
Its total-variation distance from the directly simulated $e+\mathrm{Au}$
particle-level distribution is $D_{\mathrm{TV}}^{\mathrm{pred}}=0.031$.

For the rectangular response, the condition number is the ratio of its largest
singular value to its smallest nonzero singular value,
\begin{equation}
 \kappa=\sigma_{\max}(R^p)/\sigma_{\min}(R^p),
\label{eq:supp-condition-number}
\end{equation}
with $\kappa=14.0$ for the nominal $40\times6$ matrix.

Hadronization dependence is assessed using four samples distinct from the
nominal sample above, one for each prescription in
Sec.~\ref{app:event-assembly}.  For every prescription, the perturbative event
and common PYTHIA final-state shower are fixed, and the same model is used for
both the $e+p$ response and the corresponding $e+\mathrm{Au}$ event sample.
Across the four models, the response condition numbers span $13.0$--$16.2$,
the joint $e+p$ comparisons give $D_{\mathrm{TV}}=0.0023$--$0.0045$, and the
$e+\mathrm{Au}$ $k_T$ comparisons give
$D_{\mathrm{TV}}=0.068$--$0.073$.

Without explicit jet reconstruction, these particle-level comparisons assess
the application of an $e+p$ response to $e+\mathrm{Au}$.  Detector effects are
omitted, and the response and nuclear event sample use the same hadronization
prescription.

\clearpage
\twocolumngrid

\bibliography{references}

@article{Gribov:1983ivg,
    author = "Gribov, L. V. and Levin, E. M. and Ryskin, M. G.",
    title = "{Semihard Processes in QCD}",
    doi = "10.1016/0370-1573(83)90022-4",
    journal = "Phys. Rept.",
    volume = "100",
    pages = "1--150",
    year = "1983"
}

@misc{Duan:2026jrk,
    author = "Duan, Haowu and Yi, Cong and Dai, Si-Wei and Wei, Shu-Yi and Zhao, Wenbin and Zheng, Liang",
    title = "{CGC-py: A Monte Carlo Event Generator for Gluon Saturation Physics}",
    eprint = "2609.03434",
    archivePrefix = "arXiv",
    primaryClass = "hep-ph",
    month = sep,
    year = "2026"
}

@article{Mueller:1985wy,
    author = "Mueller, Alfred H. and Qiu, Jian-wei",
    title = "{Gluon Recombination and Shadowing at Small Values of x}",
    reportNumber = "CU-TP-322",
    doi = "10.1016/0550-3213(86)90164-1",
    journal = "Nucl. Phys. B",
    volume = "268",
    pages = "427--452",
    year = "1986"
}

@article{Mueller:1989st,
    author = "Mueller, Alfred H.",
    title = "{Small x Behavior and Parton Saturation: A QCD Model}",
    reportNumber = "CU-TP-441a",
    doi = "10.1016/0550-3213(90)90173-B",
    journal = "Nucl. Phys. B",
    volume = "335",
    pages = "115--137",
    year = "1990"
}

@article{McLerran:1993ni,
    author = "McLerran, Larry D. and Venugopalan, Raju",
    title = "{Computing quark and gluon distribution functions for very large nuclei}",
    eprint = "hep-ph/9309289",
    archivePrefix = "arXiv",
    reportNumber = "TPI-MINN-93-44-T, NUC-MINN-93-24-T, HEP-UMN-TH-1220-93",
    doi = "10.1103/PhysRevD.49.2233",
    journal = "Phys. Rev. D",
    volume = "49",
    pages = "2233--2241",
    year = "1994"
}

@article{McLerran:1993ka,
    author = "McLerran, Larry D. and Venugopalan, Raju",
    title = "{Gluon distribution functions for very large nuclei at small transverse momentum}",
    eprint = "hep-ph/9311205",
    archivePrefix = "arXiv",
    reportNumber = "TPI-MINN-93-52-T, NUC-MINN-93-28-T, UMN-TH-1224-93",
    doi = "10.1103/PhysRevD.49.3352",
    journal = "Phys. Rev. D",
    volume = "49",
    pages = "3352--3355",
    year = "1994"
}

@article{McLerran:1994vd,
    author = "McLerran, Larry D. and Venugopalan, Raju",
    title = "{Green's functions in the color field of a large nucleus}",
    eprint = "hep-ph/9402335",
    archivePrefix = "arXiv",
    reportNumber = "TPI-MINN-94-7-T, NUC-MINN-94-2-T, HEP-MINN-94-1242-T",
    doi = "10.1103/PhysRevD.50.2225",
    journal = "Phys. Rev. D",
    volume = "50",
    pages = "2225--2233",
    year = "1994"
}

@article{Gelis:2010nm,
    author = "Gelis, Francois and Iancu, Edmond and Jalilian-Marian, Jamal and Venugopalan, Raju",
    title = "{The Color Glass Condensate}",
    eprint = "1002.0333",
    archivePrefix = "arXiv",
    primaryClass = "hep-ph",
    doi = "10.1146/annurev.nucl.010909.083629",
    journal = "Ann. Rev. Nucl. Part. Sci.",
    volume = "60",
    pages = "463--489",
    year = "2010"
}

@inbook{Iancu:2003xm,
    author = "Iancu, Edmond and Venugopalan, Raju",
    editor = "Hwa, Rudolph C. and Wang, Xin-Nian",
    title = "{The Color glass condensate and high-energy scattering in QCD}",
    booktitle = "{Quark-gluon plasma 4}",
    eprint = "hep-ph/0303204",
    archivePrefix = "arXiv",
    doi = "10.1142/9789812795533_0005",
    pages = "249--3363",
    month = "3",
    year = "2003"
}

@article{Accardi:2012qut,
    author = "Accardi, A. and others",
    editor = "Deshpande, A. and Meziani, Z. E. and Qiu, J. W.",
    title = "{Electron Ion Collider: The Next QCD Frontier}: {Understanding the glue that binds us all}",
    eprint = "1212.1701",
    archivePrefix = "arXiv",
    primaryClass = "nucl-ex",
    reportNumber = "BNL-98815-2012-JA, JLAB-PHY-12-1652",
    doi = "10.1140/epja/i2016-16268-9",
    journal = "Eur. Phys. J. A",
    volume = "52",
    number = "9",
    pages = "268",
    year = "2016"
}

@article{AbdulKhalek:2021gbh,
    author = "Abdul Khalek, R. and others",
    title = "{Science Requirements and Detector Concepts for the Electron-Ion Collider}: {EIC Yellow Report}",
    eprint = "2103.05419",
    archivePrefix = "arXiv",
    primaryClass = "physics.ins-det",
    reportNumber = "BNL-220990-2021-FORE, JLAB-PHY-21-3198, LA-UR-21-20953",
    doi = "10.1016/j.nuclphysa.2022.122447",
    journal = "Nucl. Phys. A",
    volume = "1026",
    pages = "122447",
    year = "2022"
}

@article{Kovchegov:1999yj,
    author = "Kovchegov, Yuri V.",
    title = "{Small x F(2) structure function of a nucleus including multiple pomeron exchanges}",
    eprint = "hep-ph/9901281",
    archivePrefix = "arXiv",
    reportNumber = "NUC-MN-99-1-T, TPI-MINN-99-05",
    doi = "10.1103/PhysRevD.60.034008",
    journal = "Phys. Rev. D",
    volume = "60",
    pages = "034008",
    year = "1999"
}

@article{Albacete:2010sy,
    author = "Albacete, Javier L. and Armesto, Nestor and Milhano, Jose Guilherme and Quiroga-Arias, Paloma and Salgado, Carlos A.",
    title = "{AAMQS: A non-linear QCD analysis of new HERA data at small-x including heavy quarks}",
    eprint = "1012.4408",
    archivePrefix = "arXiv",
    primaryClass = "hep-ph",
    doi = "10.1140/epjc/s10052-011-1705-3",
    journal = "Eur. Phys. J. C",
    volume = "71",
    pages = "1705",
    year = "2011"
}

@article{Dominguez:2011wm,
    author = "Dominguez, Fabio and Marquet, Cyrille and Xiao, Bo-Wen and Yuan, Feng",
    title = "{Universality of Unintegrated Gluon Distributions at small x}",
    eprint = "1101.0715",
    archivePrefix = "arXiv",
    primaryClass = "hep-ph",
    doi = "10.1103/PhysRevD.83.105005",
    journal = "Phys. Rev. D",
    volume = "83",
    pages = "105005",
    year = "2011"
}

@article{Dominguez:2011br,
    author = "Dominguez, Fabio and Qiu, Jian-Wei and Xiao, Bo-Wen and Yuan, Feng",
    title = "{On the linearly polarized gluon distributions in the color dipole model}",
    eprint = "1109.6293",
    archivePrefix = "arXiv",
    primaryClass = "hep-ph",
    doi = "10.1103/PhysRevD.85.045003",
    journal = "Phys. Rev. D",
    volume = "85",
    pages = "045003",
    year = "2012"
}

@article{Metz:2011wb,
    author = "Metz, Andreas and Zhou, Jian",
    title = "{Distribution of linearly polarized gluons inside a large nucleus}",
    eprint = "1105.1991",
    archivePrefix = "arXiv",
    primaryClass = "hep-ph",
    doi = "10.1103/PhysRevD.84.051503",
    journal = "Phys. Rev. D",
    volume = "84",
    pages = "051503",
    year = "2011"
}

@article{Iancu:2015joa,
    author = "Iancu, E. and Madrigal, J. D. and Mueller, A. H. and Soyez, G. and Triantafyllopoulos, D. N.",
    title = "{Collinearly-improved BK evolution meets the HERA data}",
    eprint = "1507.03651",
    archivePrefix = "arXiv",
    primaryClass = "hep-ph",
    doi = "10.1016/j.physletb.2015.09.071",
    journal = "Phys. Lett. B",
    volume = "750",
    pages = "643--652",
    year = "2015"
}

@article{Dumitru:2015gaa,
    author = "Dumitru, Adrian and Lappi, Tuomas and Skokov, Vladimir",
    title = "{Distribution of Linearly Polarized Gluons and Elliptic Azimuthal Anisotropy in Deep Inelastic Scattering Dijet Production at High Energy}",
    eprint = "1508.04438",
    archivePrefix = "arXiv",
    primaryClass = "hep-ph",
    doi = "10.1103/PhysRevLett.115.252301",
    journal = "Phys. Rev. Lett.",
    volume = "115",
    number = "25",
    pages = "252301",
    year = "2015"
}

@article{Altinoluk:2015dpi,
    author = "Altinoluk, Tolga and Armesto, N{\'e}stor and Beuf, Guillaume and Rezaeian, Amir H.",
    title = "{Diffractive Dijet Production in Deep Inelastic Scattering and Photon-Hadron Collisions in the Color Glass Condensate}",
    eprint = "1511.07452",
    archivePrefix = "arXiv",
    primaryClass = "hep-ph",
    doi = "10.1016/j.physletb.2016.05.032",
    journal = "Phys. Lett. B",
    volume = "758",
    pages = "373--383",
    year = "2016"
}

@article{Dumitru:2016jku,
    author = "Dumitru, Adrian and Skokov, Vladimir",
    title = "{$\cos(4\phi)$ azimuthal anisotropy in small-$x$ DIS dijet production beyond the leading power TMD limit}",
    eprint = "1605.02739",
    archivePrefix = "arXiv",
    primaryClass = "hep-ph",
    doi = "10.1103/PhysRevD.94.014030",
    journal = "Phys. Rev. D",
    volume = "94",
    number = "1",
    pages = "014030",
    year = "2016"
}

@article{Boer:2016fqd,
    author = {Boer, Dani{\"e}l and Mulders, Piet J. and Pisano, Cristian and Zhou, Jian},
    title = "{Asymmetries in Heavy Quark Pair and Dijet Production at an EIC}",
    eprint = "1605.07934",
    archivePrefix = "arXiv",
    primaryClass = "hep-ph",
    doi = "10.1007/JHEP08(2016)001",
    journal = "JHEP",
    volume = "08",
    pages = "001",
    year = "2016"
}

@article{Hatta:2016dxp,
    author = "Hatta, Yoshitaka and Xiao, Bo-Wen and Yuan, Feng",
    title = "{Probing the Small- x Gluon Tomography in Correlated Hard Diffractive Dijet Production in Deep Inelastic Scattering}",
    eprint = "1601.01585",
    archivePrefix = "arXiv",
    primaryClass = "hep-ph",
    reportNumber = "YITP-16-1",
    doi = "10.1103/PhysRevLett.116.202301",
    journal = "Phys. Rev. Lett.",
    volume = "116",
    number = "20",
    pages = "202301",
    year = "2016"
}

@article{Boer:2017xpy,
    author = "Boer, Daniel and Mulders, Piet J. and Zhou, Jian and Zhou, Ya-jin",
    title = "{Suppression of maximal linear gluon polarization in angular asymmetries}",
    eprint = "1702.08195",
    archivePrefix = "arXiv",
    primaryClass = "hep-ph",
    doi = "10.1007/JHEP10(2017)196",
    journal = "JHEP",
    volume = "10",
    pages = "196",
    year = "2017"
}

@article{Kotko:2017oxg,
    author = "Kotko, P. and Kutak, K. and Sapeta, S. and Stasto, A. M. and Strikman, M.",
    title = "{Estimating nonlinear effects in forward dijet production in ultra-peripheral heavy ion collisions at the LHC}",
    eprint = "1702.03063",
    archivePrefix = "arXiv",
    primaryClass = "hep-ph",
    reportNumber = "IFJPAN-IV-2017-3",
    doi = "10.1140/epjc/s10052-017-4906-6",
    journal = "Eur. Phys. J. C",
    volume = "77",
    number = "5",
    pages = "353",
    year = "2017"
}

@article{Dumitru:2018kuw,
    author = "Dumitru, Adrian and Skokov, Vladimir and Ullrich, Thomas",
    title = {{Measuring the Weizs{\"a}cker-Williams distribution of linearly polarized gluons at an electron-ion collider through dijet azimuthal asymmetries}},
    eprint = "1809.02615",
    archivePrefix = "arXiv",
    primaryClass = "hep-ph",
    doi = "10.1103/PhysRevC.99.015204",
    journal = "Phys. Rev. C",
    volume = "99",
    number = "1",
    pages = "015204",
    year = "2019"
}

@article{Mantysaari:2019csc,
    author = {M{\"a}ntysaari, Heikki and Mueller, Niklas and Schenke, Bj{\"o}rn},
    title = "{Diffractive Dijet Production and Wigner Distributions from the Color Glass Condensate}",
    eprint = "1902.05087",
    archivePrefix = "arXiv",
    primaryClass = "hep-ph",
    doi = "10.1103/PhysRevD.99.074004",
    journal = "Phys. Rev. D",
    volume = "99",
    number = "7",
    pages = "074004",
    year = "2019"
}

@article{Boussarie:2019ero,
    author = "Boussarie, R. and Grabovsky, A. V. and Szymanowski, L. and Wallon, S.",
    title = "{Towards a complete next-to-logarithmic description of forward exclusive diffractive dijet electroproduction at HERA: real corrections}",
    eprint = "1905.07371",
    archivePrefix = "arXiv",
    primaryClass = "hep-ph",
    reportNumber = "LPT-Orsay-19-22",
    doi = "10.1103/PhysRevD.100.074020",
    journal = "Phys. Rev. D",
    volume = "100",
    number = "7",
    pages = "074020",
    year = "2019"
}

@article{Salazar:2019ncp,
    author = {Salazar, Farid and Schenke, Bj{\"o}rn},
    title = "{Diffractive dijet production in impact parameter dependent saturation models}",
    eprint = "1905.03763",
    archivePrefix = "arXiv",
    primaryClass = "hep-ph",
    doi = "10.1103/PhysRevD.100.034007",
    journal = "Phys. Rev. D",
    volume = "100",
    number = "3",
    pages = "034007",
    year = "2019"
}

@article{Iancu:2020jch,
    author = "Iancu, E. and Mueller, A. H. and Triantafyllopoulos, D. N. and Wei, S. Y.",
    title = "{Saturation effects in SIDIS at very forward rapidities}",
    eprint = "2012.08562",
    archivePrefix = "arXiv",
    primaryClass = "hep-ph",
    doi = "10.1007/JHEP07(2021)196",
    journal = "JHEP",
    volume = "07",
    pages = "196",
    year = "2021"
}

@article{Bergabo:2021woe,
    author = "Bergabo, Filip and Jalilian-Marian, Jamal",
    title = "{Coherent energy loss effects in dihadron azimuthal angular correlations in Deep Inelastic Scattering at small x}",
    eprint = "2108.10428",
    archivePrefix = "arXiv",
    primaryClass = "hep-ph",
    doi = "10.1016/j.nuclphysa.2021.122358",
    journal = "Nucl. Phys. A",
    volume = "1018",
    pages = "122358",
    year = "2022"
}

@article{Boussarie:2021ybe,
    author = {Boussarie, Renaud and M{\"a}ntysaari, Heikki and Salazar, Farid and Schenke, Bj{\"o}rn},
    title = "{The importance of kinematic twists and genuine saturation effects in dijet production at the Electron-Ion Collider}",
    eprint = "2106.11301",
    archivePrefix = "arXiv",
    primaryClass = "hep-ph",
    doi = "10.1007/JHEP09(2021)178",
    journal = "JHEP",
    volume = "09",
    pages = "178",
    year = "2021"
}

@article{Zhao:2021kae,
    author = "Zhao, Ye-Yin and Xu, Ming-Mei and Chen, Li-Zhu and Zhang, Dong-Hai and Wu, Yuan-Fang",
    title = "{Suppressions of dijet azimuthal correlations in the future EIC}",
    eprint = "2105.08818",
    archivePrefix = "arXiv",
    primaryClass = "hep-ph",
    doi = "10.1103/PhysRevD.104.114032",
    journal = "Phys. Rev. D",
    volume = "104",
    number = "11",
    pages = "114032",
    year = "2021"
}

@article{Caucal:2021ent,
    author = "Caucal, Paul and Salazar, Farid and Venugopalan, Raju",
    title = "{Dijet impact factor in DIS at next-to-leading order in the Color Glass Condensate}",
    eprint = "2108.06347",
    archivePrefix = "arXiv",
    primaryClass = "hep-ph",
    doi = "10.1007/JHEP11(2021)222",
    journal = "JHEP",
    volume = "11",
    pages = "222",
    year = "2021"
}

@article{Boer:2021upt,
    author = {Boer, Dani{\"e}l and Setyadi, Chalis},
    title = "{GTMD model predictions for diffractive dijet production at EIC}",
    eprint = "2106.15148",
    archivePrefix = "arXiv",
    primaryClass = "hep-ph",
    doi = "10.1103/PhysRevD.104.074006",
    journal = "Phys. Rev. D",
    volume = "104",
    number = "7",
    pages = "074006",
    year = "2021"
}

@article{Hagiwara:2021xkf,
    author = "Hagiwara, Yoshikazu and Zhang, Cheng and Zhou, Jian and Zhou, Ya-jin",
    title = "{Probing the gluon tomography in photoproduction of dipion}",
    eprint = "2106.13466",
    archivePrefix = "arXiv",
    primaryClass = "hep-ph",
    doi = "10.1103/PhysRevD.104.094021",
    journal = "Phys. Rev. D",
    volume = "104",
    number = "9",
    pages = "094021",
    year = "2021"
}

@article{Iancu:2021rup,
    author = "Iancu, E. and Mueller, A. H. and Triantafyllopoulos, D. N.",
    title = "{Probing Parton Saturation and the Gluon Dipole via Diffractive Jet Production at the Electron-Ion Collider}",
    eprint = "2112.06353",
    archivePrefix = "arXiv",
    primaryClass = "hep-ph",
    doi = "10.1103/PhysRevLett.128.202001",
    journal = "Phys. Rev. Lett.",
    volume = "128",
    number = "20",
    pages = "202001",
    year = "2022"
}

@article{Taels:2022tza,
    author = "Taels, Pieter and Altinoluk, Tolga and Beuf, Guillaume and Marquet, Cyrille",
    title = "{Dijet photoproduction at low x at next-to-leading order and its back-to-back limit}",
    eprint = "2204.11650",
    archivePrefix = "arXiv",
    primaryClass = "hep-ph",
    doi = "10.1007/JHEP10(2022)184",
    journal = "JHEP",
    volume = "10",
    pages = "184",
    year = "2022"
}

@article{Bergabo:2022tcu,
    author = "Bergabo, Filip and Jalilian-Marian, Jamal",
    title = "{One-loop corrections to dihadron production in DIS at small x}",
    eprint = "2207.03606",
    archivePrefix = "arXiv",
    primaryClass = "hep-ph",
    doi = "10.1103/PhysRevD.106.054035",
    journal = "Phys. Rev. D",
    volume = "106",
    number = "5",
    pages = "054035",
    year = "2022"
}

@article{Bergabo:2022zhe,
    author = "Bergabo, Filip and Jalilian-Marian, Jamal",
    title = "{Single inclusive hadron production in DIS at small x: next to leading order corrections}",
    eprint = "2210.03208",
    archivePrefix = "arXiv",
    primaryClass = "hep-ph",
    doi = "10.1007/JHEP01(2023)095",
    journal = "JHEP",
    volume = "01",
    pages = "095",
    year = "2023"
}

@article{Caucal:2022ulg,
    author = {Caucal, Paul and Salazar, Farid and Schenke, Bj{\"o}rn and Venugopalan, Raju},
    title = "{Back-to-back inclusive dijets in DIS at small x: Sudakov suppression and gluon saturation at NLO}",
    eprint = "2208.13872",
    archivePrefix = "arXiv",
    primaryClass = "hep-ph",
    doi = "10.1007/JHEP11(2022)169",
    journal = "JHEP",
    volume = "11",
    pages = "169",
    year = "2022"
}

@article{Tong:2022zwp,
    author = "Tong, Xuan-Bo and Xiao, Bo-Wen and Zhang, Yuan-Yuan",
    title = "{Harmonics of Parton Saturation in Lepton-Jet Correlations at the Electron-Ion Collider}",
    eprint = "2211.01647",
    archivePrefix = "arXiv",
    primaryClass = "hep-ph",
    doi = "10.1103/PhysRevLett.130.151902",
    journal = "Phys. Rev. Lett.",
    volume = "130",
    number = "15",
    pages = "151902",
    year = "2023"
}

@article{Iancu:2022lcw,
    author = "Iancu, E. and Mueller, A. H. and Triantafyllopoulos, D. N. and Wei, S. Y.",
    title = "{Gluon dipole factorisation for diffractive dijets}",
    eprint = "2207.06268",
    archivePrefix = "arXiv",
    primaryClass = "hep-ph",
    doi = "10.1007/JHEP10(2022)103",
    journal = "JHEP",
    volume = "10",
    pages = "103",
    year = "2022"
}

@article{Hatta:2022lzj,
    author = "Hatta, Yoshitaka and Xiao, Bo-Wen and Yuan, Feng",
    title = "{Semi-inclusive diffractive deep inelastic scattering at small x}",
    eprint = "2205.08060",
    archivePrefix = "arXiv",
    primaryClass = "hep-ph",
    doi = "10.1103/PhysRevD.106.094015",
    journal = "Phys. Rev. D",
    volume = "106",
    number = "9",
    pages = "094015",
    year = "2022"
}

@article{Bergabo:2023wed,
    author = "Bergabo, Filip and Jalilian-Marian, Jamal",
    title = "{Dihadron production in DIS at small x at next-to-leading order: Transverse photons}",
    eprint = "2301.03117",
    archivePrefix = "arXiv",
    primaryClass = "hep-ph",
    doi = "10.1103/PhysRevD.107.054036",
    journal = "Phys. Rev. D",
    volume = "107",
    number = "5",
    pages = "054036",
    year = "2023"
}

@article{Tong:2023bus,
    author = "Tong, Xuan-Bo and Xiao, Bo-Wen and Zhang, Yuan-Yuan",
    title = "{Harmonics of lepton-jet correlations in inclusive and diffractive scatterings}",
    eprint = "2310.20662",
    archivePrefix = "arXiv",
    primaryClass = "hep-ph",
    doi = "10.1103/PhysRevD.109.054004",
    journal = "Phys. Rev. D",
    volume = "109",
    number = "5",
    pages = "054004",
    year = "2024"
}

@article{Rodriguez-Aguilar:2023ihz,
    author = "Rodriguez-Aguilar, Benjamin and Triantafyllopoulos, D. N. and Wei, S. Y.",
    title = "{Incoherent diffractive dijet production in electron DIS off nuclei at high energy}",
    eprint = "2302.01106",
    archivePrefix = "arXiv",
    primaryClass = "hep-ph",
    doi = "10.1103/PhysRevD.107.114007",
    journal = "Phys. Rev. D",
    volume = "107",
    number = "11",
    pages = "114007",
    year = "2023"
}

@article{Caucal:2023nci,
    author = {Caucal, Paul and Salazar, Farid and Schenke, Bj{\"o}rn and Stebel, Tomasz and Venugopalan, Raju},
    title = {{Back-to-back inclusive dijets in DIS at small x: gluon Weizs{\"a}cker-Williams distribution at NLO}},
    eprint = "2304.03304",
    archivePrefix = "arXiv",
    primaryClass = "hep-ph",
    doi = "10.1007/JHEP08(2023)062",
    journal = "JHEP",
    volume = "08",
    pages = "062",
    year = "2023"
}

@article{Caucal:2023fsf,
    author = {Caucal, Paul and Salazar, Farid and Schenke, Bj{\"o}rn and Stebel, Tomasz and Venugopalan, Raju},
    title = "{Back-to-Back Inclusive Dijets in Deep Inelastic Scattering at Small x: Complete NLO Results and Predictions}",
    eprint = "2308.00022",
    archivePrefix = "arXiv",
    primaryClass = "hep-ph",
    doi = "10.1103/PhysRevLett.132.081902",
    journal = "Phys. Rev. Lett.",
    volume = "132",
    number = "8",
    pages = "081902",
    year = "2024"
}

@article{Shao:2024nor,
    author = "Shao, Ding Yu and Shi, Yu and Zhang, Cheng and Zhou, Jian and Zhou, Ya-jin",
    title = "{Revisiting azimuthal angular asymmetries in diffractive di-jet production}",
    eprint = "2402.05465",
    archivePrefix = "arXiv",
    primaryClass = "hep-ph",
    doi = "10.1007/JHEP07(2024)189",
    journal = "JHEP",
    volume = "07",
    pages = "189",
    year = "2024"
}

@article{Altinoluk:2024vgg,
    author = "Altinoluk, Tolga and Jalilian-Marian, Jamal and Marquet, Cyrille",
    title = "{Sudakov double logs in single-inclusive hadron production in DIS at small x from the color glass condensate formalism}",
    eprint = "2406.08277",
    archivePrefix = "arXiv",
    primaryClass = "hep-ph",
    doi = "10.1103/PhysRevD.110.094056",
    journal = "Phys. Rev. D",
    volume = "110",
    number = "9",
    pages = "094056",
    year = "2024"
}

@article{Caucal:2024nsb,
    author = "Caucal, Paul and Salazar, Farid",
    title = "{Dihadron correlations in small-x DIS at NLO: transverse momentum dependent fragmentation}",
    eprint = "2405.19404",
    archivePrefix = "arXiv",
    primaryClass = "hep-ph",
    reportNumber = "INT-PUB-24-022",
    doi = "10.1007/JHEP12(2024)130",
    journal = "JHEP",
    volume = "12",
    pages = "130",
    year = "2024"
}

@article{Altinoluk:2025dwd,
    author = "Altinoluk, Tolga and Bergabo, Filip and Jalilian-Marian, Jamal and Marquet, Cyrille and Shi, Yu",
    title = "{SIDIS at small x at next-to-leading order: Transverse photon}",
    eprint = "2505.04557",
    archivePrefix = "arXiv",
    primaryClass = "hep-ph",
    doi = "10.1103/s561-vqh8",
    journal = "Phys. Rev. D",
    volume = "112",
    number = "5",
    pages = "054020",
    year = "2025"
}

@article{Caucal:2025qjg,
    author = "Caucal, Paul and Salazar, Farid",
    title = "{Small-x Factorization in the Target Fragmentation Region}",
    eprint = "2502.02634",
    archivePrefix = "arXiv",
    primaryClass = "hep-ph",
    doi = "10.1103/3b6j-th2m",
    journal = "Phys. Rev. Lett.",
    volume = "136",
    number = "8",
    pages = "081901",
    year = "2026"
}

@article{Marquet:2025jdr,
    author = "Marquet, Cyrille and Shi, Yu and Xiao, Bo-Wen",
    title = "{Unified Resummation of Soft Gluon Radiation in Heavy Meson Pair Photoproduction}",
    eprint = "2510.18949",
    archivePrefix = "arXiv",
    primaryClass = "hep-ph",
    journal = {},
    month = "10",
    year = "2025"
}

@article{Shao:2026doo,
    author = "Shao, Ding Yu and Shi, Yu and Zhang, Cheng and Zhou, Jian and Zhou, Ya-jin",
    title = "{Azimuthal decorrelation in diffractive dijet production}",
    eprint = "2606.02230",
    archivePrefix = "arXiv",
    primaryClass = "hep-ph",
    journal = {},
    month = "6",
    year = "2026"
}

@article{Marquet:2007vb,
    author = "Marquet, Cyrille",
    title = "{Forward inclusive dijet production and azimuthal correlations in p(A) collisions}",
    eprint = "0708.0231",
    archivePrefix = "arXiv",
    primaryClass = "hep-ph",
    doi = "10.1016/j.nuclphysa.2007.09.001",
    journal = "Nucl. Phys. A",
    volume = "796",
    pages = "41--60",
    year = "2007"
}

@article{Stasto:2018rci,
    author = "Stasto, Anna and Wei, Shu-Yi and Xiao, Bo-Wen and Yuan, Feng",
    title = "{On the Dihadron Angular Correlations in Forward $pA$ collisions}",
    eprint = "1805.05712",
    archivePrefix = "arXiv",
    primaryClass = "hep-ph",
    doi = "10.1016/j.physletb.2018.08.011",
    journal = "Phys. Lett. B",
    volume = "784",
    pages = "301--306",
    year = "2018"
}

@article{Mueller:2012uf,
    author = "Mueller, A. H. and Xiao, Bo-Wen and Yuan, Feng",
    title = "{Sudakov Resummation in Small-$x$ Saturation Formalism}",
    eprint = "1210.5792",
    archivePrefix = "arXiv",
    primaryClass = "hep-ph",
    doi = "10.1103/PhysRevLett.110.082301",
    journal = "Phys. Rev. Lett.",
    volume = "110",
    number = "8",
    pages = "082301",
    year = "2013"
}

@article{Mueller:2013wwa,
    author = "Mueller, A. H. and Xiao, Bo-Wen and Yuan, Feng",
    title = "{Sudakov double logarithms resummation in hard processes in the small-x saturation formalism}",
    eprint = "1308.2993",
    archivePrefix = "arXiv",
    primaryClass = "hep-ph",
    doi = "10.1103/PhysRevD.88.114010",
    journal = "Phys. Rev. D",
    volume = "88",
    number = "11",
    pages = "114010",
    year = "2013"
}

@article{Zheng:2014vka,
    author = "Zheng, L. and Aschenauer, E. C. and Lee, J. H. and Xiao, Bo-Wen",
    title = "{Probing Gluon Saturation through Dihadron Correlations at an Electron-Ion Collider}",
    eprint = "1403.2413",
    archivePrefix = "arXiv",
    primaryClass = "hep-ph",
    doi = "10.1103/PhysRevD.89.074037",
    journal = "Phys. Rev. D",
    volume = "89",
    number = "7",
    pages = "074037",
    year = "2014"
}

@article{Akcakaya:2012si,
    author = {Akcakaya, Emin and Sch{\"a}fer, Andreas and Zhou, Jian},
    title = "{Azimuthal asymmetries for quark pair production in pA collisions}",
    eprint = "1208.4965",
    archivePrefix = "arXiv",
    primaryClass = "hep-ph",
    doi = "10.1103/PhysRevD.87.054010",
    journal = "Phys. Rev. D",
    volume = "87",
    number = "5",
    pages = "054010",
    year = "2013"
}

@article{Kotko:2015ura,
    author = "Kotko, P. and Kutak, K. and Marquet, C. and Petreska, E. and Sapeta, S. and van Hameren, A.",
    title = "{Improved TMD factorization for forward dijet production in dilute-dense hadronic collisions}",
    eprint = "1503.03421",
    archivePrefix = "arXiv",
    primaryClass = "hep-ph",
    reportNumber = "CERN-PH-TH-2015-045, CPHT-RR005.0315, IFJPAN-IV-2015-2",
    doi = "10.1007/JHEP09(2015)106",
    journal = "JHEP",
    volume = "09",
    pages = "106",
    year = "2015"
}

@article{vanHameren:2014ala,
    author = "van Hameren, A. and Kotko, P. and Kutak, K. and Sapeta, S.",
    title = "{Small-$x$ dynamics in forward-central dijet decorrelations at the LHC}",
    eprint = "1404.6204",
    archivePrefix = "arXiv",
    primaryClass = "hep-ph",
    reportNumber = "IFJPAN-IV-2014-6, CERN-PH-TH-2014-070",
    doi = "10.1016/j.physletb.2014.09.005",
    journal = "Phys. Lett. B",
    volume = "737",
    pages = "335--340",
    year = "2014"
}

@article{vanHameren:2016ftb,
    author = "van Hameren, A. and Kotko, P. and Kutak, K. and Marquet, C. and Petreska, E. and Sapeta, S.",
    title = "{Forward di-jet production in p+Pb collisions in the small-x improved TMD factorization framework}",
    eprint = "1607.03121",
    archivePrefix = "arXiv",
    primaryClass = "hep-ph",
    reportNumber = "CERN-TH-2016-159, CPHT-RR036.072016, IFJPAN-IV-2016-X, CERN-TH-2016-XXX",
    doi = "10.1007/JHEP12(2016)034",
    journal = "JHEP",
    volume = "12",
    pages = "034",
    year = "2016",
    note = "[Erratum: JHEP 02, 158 (2019)]"
}

@article{vanHameren:2019ysa,
    author = "van Hameren, Andreas and Kotko, Piotr and Kutak, Krzysztof and Sapeta, Sebastian",
    title = "{Broadening and saturation effects in dijet azimuthal correlations in p-p and p-Pb collisions at $\mathbf{\sqrt{s}} = $ 5.02 TeV}",
    eprint = "1903.01361",
    archivePrefix = "arXiv",
    primaryClass = "hep-ph",
    reportNumber = "IFJPAN-IV-2019-3",
    doi = "10.1016/j.physletb.2019.06.055",
    journal = "Phys. Lett. B",
    volume = "795",
    pages = "511--515",
    year = "2019"
}

@article{vanHameren:2020rqt,
    author = "van Hameren, A. and Kotko, P. and Kutak, K. and Sapeta, S.",
    title = "{Sudakov effects in central-forward dijet production in high energy factorization}",
    eprint = "2010.13066",
    archivePrefix = "arXiv",
    primaryClass = "hep-ph",
    reportNumber = "IFJPAN-IV-2020-8",
    doi = "10.1016/j.physletb.2021.136078",
    journal = "Phys. Lett. B",
    volume = "814",
    pages = "136078",
    year = "2021"
}

@article{Al-Mashad:2022zbq,
    author = "Al-Mashad, M. Abdullah and van Hameren, A. and Kakkad, H. and Kotko, P. and Kutak, K. and van Mechelen, P. and Sapeta, S.",
    title = "{Dijet azimuthal correlations in p-p and p-Pb collisions at forward LHC calorimeters}",
    eprint = "2210.06613",
    archivePrefix = "arXiv",
    primaryClass = "hep-ph",
    doi = "10.1007/JHEP12(2022)131",
    journal = "JHEP",
    volume = "12",
    pages = "131",
    year = "2022"
}

@article{Gao:2026azd,
    author = "Gao, Zhan and Marquet, Cyrille and Shi, Yu and Xiao, Bo-Wen",
    title = "{Probing Saturation Effect in Heavy Meson Pair Correlation in Forward $pA$ Collisions}",
    eprint = "2605.01527",
    archivePrefix = "arXiv",
    primaryClass = "hep-ph",
    journal = {},
    month = "5",
    year = "2026"
}

@article{Xiao:2017yya,
    author = "Xiao, Bo-Wen and Yuan, Feng and Zhou, Jian",
    title = "{Transverse Momentum Dependent Parton Distributions at Small-x}",
    eprint = "1703.06163",
    archivePrefix = "arXiv",
    primaryClass = "hep-ph",
    doi = "10.1016/j.nuclphysb.2017.05.012",
    journal = "Nucl. Phys. B",
    volume = "921",
    pages = "104--126",
    year = "2017"
}

@article{Zhou:2016tfe,
    author = "Zhou, Jian",
    title = "{The evolution of the small x gluon TMD}",
    eprint = "1603.07426",
    archivePrefix = "arXiv",
    primaryClass = "hep-ph",
    doi = "10.1007/JHEP06(2016)151",
    journal = "JHEP",
    volume = "06",
    pages = "151",
    year = "2016"
}

@article{Zhou:2018lfq,
    author = "Zhou, Jian",
    title = "{Scale dependence of the small x transverse momentum dependent gluon distribution}",
    eprint = "1807.00506",
    archivePrefix = "arXiv",
    primaryClass = "hep-ph",
    doi = "10.1103/PhysRevD.99.054026",
    journal = "Phys. Rev. D",
    volume = "99",
    number = "5",
    pages = "054026",
    year = "2019"
}

@article{Dominguez:2011gc,
    author = "Dominguez, Fabio and Mueller, A. H. and Munier, St{\'e}phane and Xiao, Bo-Wen",
    title = {{On the small-$x$ evolution of the color quadrupole and the Weizs{\"a}cker{\textendash}Williams gluon distribution}},
    eprint = "1108.1752",
    archivePrefix = "arXiv",
    primaryClass = "hep-ph",
    doi = "10.1016/j.physletb.2011.09.104",
    journal = "Phys. Lett. B",
    volume = "705",
    pages = "106--111",
    year = "2011"
}

@article{Beuf:2014uia,
    author = "Beuf, Guillaume",
    title = "{Improving the kinematics for low-$x$ QCD evolution equations in coordinate space}",
    eprint = "1401.0313",
    archivePrefix = "arXiv",
    primaryClass = "hep-ph",
    doi = "10.1103/PhysRevD.89.074039",
    journal = "Phys. Rev. D",
    volume = "89",
    number = "7",
    pages = "074039",
    year = "2014"
}

@article{Liu:2022xsc,
    author = "Liu, Hao-yu and Liu, Xiao-hui and Shi, Yu and Zheng, Du-xin and Zhou, Jian",
    title = "{Kinematic constraint in the BFKL evolution near threshold region}",
    eprint = "2204.00262",
    archivePrefix = "arXiv",
    primaryClass = "hep-ph",
    doi = "10.1103/PhysRevD.106.036026",
    journal = "Phys. Rev. D",
    volume = "106",
    number = "3",
    pages = "036026",
    year = "2022"
}

@article{Zheng:2019zul,
    author = "Zheng, Du-Xin and Zhou, Jian",
    title = "{Sudakov suppression of the Balitsky-Kovchegov kernel}",
    eprint = "1906.06825",
    archivePrefix = "arXiv",
    primaryClass = "hep-ph",
    doi = "10.1007/JHEP11(2019)177",
    journal = "JHEP",
    volume = "11",
    pages = "177",
    year = "2019"
}

@article{Marquet:2005zf,
    author = "Marquet, C. and Soyez, G.",
    title = "{The Balitsky-Kovchegov equation in full momentum space}",
    eprint = "hep-ph/0504080",
    archivePrefix = "arXiv",
    doi = "10.1016/j.nuclphysa.2005.05.198",
    journal = "Nucl. Phys. A",
    volume = "760",
    pages = "208--222",
    year = "2005"
}

@article{Liu:2023aqb,
    author = "Liu, Hao-Yu and Liu, Xiaohui and Pan, Ji-Chen and Yuan, Feng and Zhu, Hua Xing",
    title = "{Nucleon Energy Correlators for the Color Glass Condensate}",
    eprint = "2301.01788",
    archivePrefix = "arXiv",
    primaryClass = "hep-ph",
    doi = "10.1103/PhysRevLett.130.181901",
    journal = "Phys. Rev. Lett.",
    volume = "130",
    number = "18",
    pages = "181901",
    year = "2023"
}

@article{Mantysaari:2025mht,
    author = {M{\"a}ntysaari, Heikki and Tawabutr, Yossathorn and Tong, Xuan-Bo},
    title = "{Nucleon energy correlators for the odderon}",
    eprint = "2503.20157",
    archivePrefix = "arXiv",
    primaryClass = "hep-ph",
    doi = "10.1103/rmzx-tgm2",
    journal = "Phys. Rev. D",
    volume = "112",
    number = "11",
    pages = "114027",
    year = "2025"
}

@article{Kang:2026hig,
    author = "Kang, Zhong-Bo and Kao, Robert and Li, Meijian and Penttala, Jani",
    title = "{One-point energy correlator for deep inelastic scattering at small $x$}",
    eprint = "2603.02300",
    archivePrefix = "arXiv",
    primaryClass = "hep-ph",
    journal = {},
    month = "3",
    year = "2026"
}

@article{Mantysaari:2026zte,
    author = {M{\"a}ntysaari, Heikki and Shi, Yu and Tawabutr, Yossathorn and Tong, Xuan-Bo},
    title = "{Gluonic nucleon energy correlators and fracture functions for Color Glass Condensate}",
    eprint = "2608.10955",
    archivePrefix = "arXiv",
    primaryClass = "hep-ph",
    journal = {},
    month = "8",
    year = "2026"
}

@article{Liu:2022wop,
    author = "Liu, Xiaohui and Zhu, Hua Xing",
    title = "{Nucleon Energy Correlators}",
    eprint = "2209.02080",
    archivePrefix = "arXiv",
    primaryClass = "hep-ph",
    doi = "10.1103/PhysRevLett.130.091901",
    journal = "Phys. Rev. Lett.",
    volume = "130",
    number = "9",
    pages = "091901",
    year = "2023"
}

@article{Ke:2023xeo,
    author = "Ke, Weiyao and Zhang, Yuan-Yuan and Xing, Hongxi and Wang, Xin-Nian",
    title = "{Event generator for jet tomography in electron-ion collisions}",
    eprint = "2304.10779",
    archivePrefix = "arXiv",
    primaryClass = "hep-ph",
    reportNumber = "LA-UR-23-24141",
    doi = "10.1103/PhysRevD.110.034001",
    journal = "Phys. Rev. D",
    volume = "110",
    number = "3",
    pages = "034001",
    year = "2024"
}

@article{Shi:2022hee,
    author = "Shi, Yu and Wei, Shu-Yi and Zhou, Jian",
    title = "{Parton shower generator based on the Gribov-Levin-Ryskin equation}",
    eprint = "2211.07174",
    archivePrefix = "arXiv",
    primaryClass = "hep-ph",
    doi = "10.1103/PhysRevD.107.016017",
    journal = "Phys. Rev. D",
    volume = "107",
    number = "1",
    pages = "016017",
    year = "2023"
}

@article{Shi:2023ejp,
    author = "Shi, Yu and Wei, Shu-Yi and Zhou, Jian",
    title = "{Parton shower algorithm with the saturation effect}",
    eprint = "2307.04185",
    archivePrefix = "arXiv",
    primaryClass = "hep-ph",
    doi = "10.1103/PhysRevD.108.096025",
    journal = "Phys. Rev. D",
    volume = "108",
    number = "9",
    pages = "096025",
    year = "2023"
}

@article{Kutak:2011fu,
    author = "Kutak, Krzysztof and Golec-Biernat, Krzysztof and Jadach, Stanislaw and Skrzypek, Maciej",
    title = "{Nonlinear equation for coherent gluon emission}",
    eprint = "1111.6928",
    archivePrefix = "arXiv",
    primaryClass = "hep-ph",
    doi = "10.1007/JHEP02(2012)117",
    journal = "JHEP",
    volume = "02",
    pages = "117",
    year = "2012"
}

@article{Jung:2000hk,
    author = "Jung, H. and Salam, G. P.",
    title = "{Hadronic final state predictions from CCFM: The Hadron level Monte Carlo generator CASCADE}",
    eprint = "hep-ph/0012143",
    archivePrefix = "arXiv",
    reportNumber = "CERN-TH-2000-318, DESY-00-151, LUNFD6-NFFL-7189-2000, LPTHE-00-43",
    doi = "10.1007/s100520100604",
    journal = "Eur. Phys. J. C",
    volume = "19",
    pages = "351--360",
    year = "2001"
}

@article{Baranov:2021uol,
    author = "Baranov, S. and others",
    collaboration = "CASCADE",
    title = "{CASCADE3 A Monte Carlo event generator based on TMDs}",
    eprint = "2101.10221",
    archivePrefix = "arXiv",
    primaryClass = "hep-ph",
    reportNumber = "DESY-21-005",
    doi = "10.1140/epjc/s10052-021-09203-8",
    journal = "Eur. Phys. J. C",
    volume = "81",
    number = "5",
    pages = "425",
    year = "2021"
}

@article{Hautmann:2022xuc,
    author = "Hautmann, F. and Hentschinski, M. and Keersmaekers, L. and Kusina, A. and Kutak, K. and Lelek, A.",
    title = "{A parton branching with transverse momentum dependent splitting functions}",
    eprint = "2205.15873",
    archivePrefix = "arXiv",
    primaryClass = "hep-ph",
    reportNumber = "CERN-TH-2022-087, IFJPAN-IV-2022-8",
    doi = "10.1016/j.physletb.2022.137276",
    journal = "Phys. Lett. B",
    volume = "833",
    pages = "137276",
    year = "2022"
}

@article{Lipatov:2023ypn,
    author = "Lipatov, A. V. and Malyshev, M. A.",
    title = "{TMD parton showers for associated {\ensuremath{\gamma}}+jet production in electron-proton collisions at high energies}",
    eprint = "2305.04005",
    archivePrefix = "arXiv",
    primaryClass = "hep-ph",
    doi = "10.1103/PhysRevD.108.014022",
    journal = "Phys. Rev. D",
    volume = "108",
    number = "1",
    pages = "014022",
    year = "2023"
}

@article{Bierlich:2022pfr,
    author = "Bierlich, Christian and others",
    title = "{A comprehensive guide to the physics and usage of PYTHIA 8.3}",
    eprint = "2203.11601",
    archivePrefix = "arXiv",
    primaryClass = "hep-ph",
    reportNumber = "LU-TP 22-16, MCNET-22-04, FERMILAB-PUB-22-227-SCD",
    doi = "10.21468/SciPostPhysCodeb.8",
    journal = "SciPost Phys. Codeb.",
    volume = "2022",
    pages = "8",
    year = "2022"
}

@article{DAgostini:1994fjx,
    author = "D'Agostini, G.",
    title = "{A Multidimensional unfolding method based on Bayes' theorem}",
    reportNumber = "DESY-94-099",
    doi = "10.1016/0168-9002(95)00274-X",
    journal = "Nucl. Instrum. Meth. A",
    volume = "362",
    pages = "487--498",
    year = "1995"
}

@article{Bellm:2015jjp,
    author = "Bellm, Johannes and others",
    title = "{Herwig 7.0/Herwig++ 3.0 release note}",
    eprint = "1512.01178",
    archivePrefix = "arXiv",
    primaryClass = "hep-ph",
    reportNumber = "CERN-PH-TH-2015-289, MAN-HEP-2015-15, IFJPAN-IV-2015-13, KA-TP-18-2015, DCPT-15-142, MCNET-15-28, IPPP-15-71, HERWIG-2015-01",
    doi = "10.1140/epjc/s10052-016-4018-8",
    journal = "Eur. Phys. J. C",
    volume = "76",
    number = "4",
    pages = "196",
    year = "2016"
}

@article{Winter:2003tt,
    author = "Winter, Jan-Christopher and Krauss, Frank and Soff, Gerhard",
    title = "{A Modified cluster hadronization model}",
    eprint = "hep-ph/0311085",
    archivePrefix = "arXiv",
    reportNumber = "CERN-TH-2003-272",
    doi = "10.1140/epjc/s2004-01960-8",
    journal = "Eur. Phys. J. C",
    volume = "36",
    pages = "381--395",
    year = "2004"
}

@article{JETSCAPE:2023ewn,
    author = "Angerami, Aaron and others",
    collaboration = "JETSCAPE",
    title = "{Hybrid Hadronization of Jet Showers from $e^++e^-$ to $A+A$ with JETSCAPE}",
    eprint = "2310.20631",
    archivePrefix = "arXiv",
    primaryClass = "hep-ph",
    doi = "10.22323/1.438.0166",
    journal = "PoS",
    volume = "HardProbes2023",
    pages = "166",
    year = "2024"
}

@article{Motyka:2023pmt,
    author = "Motyka, Leszek and Sadzikowski, Mariusz",
    title = "{Twist decomposition of non-linear effects in Balitsky{\textendash}Kovchegov evolution of proton structure functions}",
    eprint = "2306.02118",
    archivePrefix = "arXiv",
    primaryClass = "hep-ph",
    doi = "10.1140/epjc/s10052-023-12241-z",
    journal = "Eur. Phys. J. C",
    volume = "83",
    number = "11",
    pages = "1062",
    year = "2023"
}

@article{Zhang:2021tcc,
    author = "Zhang, Yuan-Yuan and Wang, Xin-Nian",
    title = "{Parton rescattering and gluon saturation in dijet production at EIC}",
    eprint = "2104.04520",
    archivePrefix = "arXiv",
    primaryClass = "hep-ph",
    doi = "10.1103/PhysRevD.105.034015",
    journal = "Phys. Rev. D",
    volume = "105",
    number = "3",
    pages = "034015",
    year = "2022"
}

@article{Shi:2021hwx,
    author = "Shi, Yu and Wang, Lei and Wei, Shu-Yi and Xiao, Bo-Wen",
    title = "{Pursuing the Precision Study for Color Glass Condensate in Forward Hadron Productions}",
    eprint = "2112.06975",
    archivePrefix = "arXiv",
    primaryClass = "hep-ph",
    doi = "10.1103/PhysRevLett.128.202302",
    journal = "Phys. Rev. Lett.",
    volume = "128",
    number = "20",
    pages = "202302",
    year = "2022"
}

\end{document}